\documentclass[twocolumn]{aastex631} 
\usepackage{csvsimple}
\usepackage{multirow,bigdelim}

\usepackage{amsmath}
\usepackage{mathtools}
\usepackage{hyperref}

\newcommand{\sigSB}{\sigma_{\rm SB}}
\newcommand{\T}{{\rm T}}
\newcommand{\msun}{\mbox{$\,{\rm M}_\odot$}} 
\newcommand{\lsun}{\mbox{$\,{\rm L}_\odot$}}

\newcommand{\teff}{{\rm T_{eff}}}

\newcommand{\om}{`Oumuamua}

\usepackage{colortbl}
\definecolor{tablegray}{rgb}{0.89, 0.89, 0.89}
\shorttitle{Exo-Comets from Post-MS Stars}
\shortauthors{Levine et al.}

\begin{document}

\title{Interstellar Comets from Post--Main Sequence Systems as Tracers of Extrasolar Oort Clouds}
\correspondingauthor{W. Garrett Levine}

\email{garrett.levine@yale.edu}

\author[0000-0002-1422-4430]{W. Garrett Levine}
\affil{Dept. of Astronomy, Yale University. New Haven, CT 06511, USA}

\author[0000-0002-0140-4475]{Aster G. Taylor}
\affil{Dept. of Astronomy \& Astrophysics, University of Chicago, Chicago, IL 60637, USA}

\author[0000-0002-0726-6480]{Darryl Z. Seligman}
\affiliation{Dept. of Astronomy \& Carl Sagan Institute, Cornell University, 122 Sciences Drive, Ithaca, NY, 14853, USA}

\author[0000-0002-7783-6397]{Devin J. Hoover}
\affil{Dept. of Astronomy \& Astrophysics, University of Chicago, Chicago, IL 60637, USA}


\author[0000-0001-7830-028X]{Robert Jedicke}
\affil{Institute for Astronomy, University of Hawaii, 2680 Woodlawn Dr, Honolulu, HI 96822, USA}

\author[0000-0002-8716-0482]{Jennifer B. Bergner}
\affil{Dept. of Chemistry, University of California, Berkeley, CA, 94720, USA}

\author[0000-0002-3253-2621]{Gregory P. Laughlin}
\affil{Dept. of Astronomy, Yale University. New Haven, CT 06511, USA}

\begin{abstract}
Interstellar small bodies are unique probes into the histories of exoplanetary systems. One hypothesized class of interlopers are ``Jurads," exo-comets released into the Milky Way during the post--main sequence as the thermally-pulsing asymptotic giant branch (AGB) host stars lose mass. In this study, we assess the prospects for the Legacy Survey of Space and Time (LSST) to detect a Jurad and examine whether such an interloper would be observationally distinguishable from exo-comets ejected during the (pre--)main sequence. Using analytic and numerical methods, we estimate the fraction of exo-Oort Cloud objects that are released from $1-8\,\msun$ stars during post--main sequence evolution. We quantify the extent to which small bodies are altered by the increased luminosity and stellar outflows during the AGB, finding that some Jurads may lack hypervolatiles and that stellar winds could deposit dust that covers the entire exo-comet surface. Next, we construct models of the interstellar small body reservoir for various size-frequency distributions and examine the LSST's ability to detect members of those hypothesized populations. Combining these analyses, we highlight the joint constraints that the LSST will place on power-law size-frequency distribution slopes, characteristic sizes, and the total mass sequestered in the minor planets of exo--Oort Clouds. Even with the LSST's increased search volume compared to contemporary surveys, we find that detecting a Jurad is unlikely but not infeasible given the current understanding of (exo)planet formation.
\vspace{10mm}
\end{abstract}


\section{Introduction} \label{sec:intro}

There exists an established scientific precedent of using the orbital distributions and compositions of minor planets to infer the dynamical evolution of the solar system. Small bodies detected in wide-field surveys \citep{jewitt1993kuiper}, targeted follow-up observations \citep{schwamb2019colossos}, and theoretical studies \citep{fernandez1984migrationNeptune, Hahn99, Gomes2004, Nesvorny2018} have together revealed that the giant planets experienced an epoch of migration and/or orbital instability several $10^{8}$ yr after their initial formation. A preponderance of the evidence, such as the existence of the Kuiper Belt, supports a scenario where the solar system contained approximately $30\,\text{M}_{\oplus}$ of planetesimals after the main epoch of planet formation, most of which was subsequently ejected into the interstellar medium (ISM) by the giant planets \citep{Hahn99,Tsiganis2005,Levison2008}. A small amount ($\sim$0.1 M$_{\oplus}$) was scattered into the Kuiper Belt \citep{gladman2001kuiperBelt}. A modest, less constrained mass ($\sim$2M$_{\oplus}$) was scattered into the Oort Cloud \citep{Oort1950, weissman1983oort, weissman1990oort, Dones15}.

Likewise, interstellar small bodies can illuminate extrasolar environments. Since 2017, the first two interlopers have provided close-up glimpses of extrasolar ejecta. The origin of 1I/\om{} \citep{Williams17,Meech2017} has been fiercely debated, but each of the hypotheses on its bulk composition points towards previously unconsidered astrophysics in exoplanetary systems \citep{Jewitt2022, seligman2023reviewMoroMartin, fitzsimmons2023review}. Recently, \cite{bergner2023H2} demonstrated that `Oumuamua's properties were consistent with an amorphous water ice comet with radiolitically-produced H$_2$ via galactic cosmic rays. 2I/Borisov, in contrast, was promptly identified as an exo-comet from its extended coma \citep{Jewitt2019,Guzik2020}. The CO-dominated outgassing \citep{Cordiner2020,Bodewits2020,Yang2021} was strikingly different from typical H$_{2}$O-dominated solar system comets and hinted that Borisov's formation might have occurred at lower temperature than solar system analogs \citep{Lisse2022, seligman2022borisovCO}. \citet{Bodewits2020} ascribed this peculiar composition to formation in an M-dwarf system, while \citet{Cordiner2020} suggested that Borisov may have formed beyond the CO ice line of a Solar-like star.

In the upcoming years, the Legacy Survey of Space \& Time (``the LSST") will be executed at the Vera Rubin Observatory (Rubin) and produce an unparalleled census of small bodies in the solar system \citep{jones2009LSSTSolarSystem}. Due to the ``wide-fast-deep" strategy \citep{ivezic2019lsst}, the LSST should also discover exo-comets passing through the solar system \citep{moro2009will,cook2016realistic, engelhardt2017observational,hoover2022population}. Although only of order $10$ (albeit with many uncertainties) small bodies of interstellar origin are expected to be identified \citep{hoover2022population}, the physical, kinematic, and dynamical properties of exo-comets will nonetheless provide critical knowledge towards comparing the Sun's planetary system to its peers \citep{Jewitt2022}.

Because the Sun's interstellar ejecta was generated during the main sequence, research on extrasolar small bodies has mostly considered the prevalence of analogous objects \citep{sekanina1976probability, mcglynn1989nondetection, francis2005, moro2009will, engelhardt2017observational}. However, small bodies should also be efficiently ejected from Oort Cloud-like semimajor axes during the asymptotic giant branch (AGB) phase of the lifetime of a star \citep{veras2011escape1, hansen2017postMS, katz2018interstellar, Rafikov2018b, moro2018originII}. The stellar envelope is rapidly lost during this phase via thermal pulses, and the shrinking gravitational potential well perturbs companions within the system and releases some into interstellar space. Indeed, recent observations of solids accreting onto white dwarfs \citep{Xu2013,Farihi2013,Xu2014,Wilson2015,Wilson2016,kaiser2021wdAccretionLithium} along with transits of short-period planets and planetesimals orbiting white dwarfs \citep{vanderburg2015WDplanetesimal, manser2019wdPlanetesimal, vanderburg2020WDplanet} confirm that post--main sequence planetary systems are dynamically active environments.

Upon the arrival of \om, many authors \citep{hansen2017postMS, Rafikov2018b, katz2018interstellar, moro2018originII}, considered whether this interloper could have been ejected during its host star's post--main sequence. Although \om's physical properties are unparalleled among the solar system's minor planets, the inbound kinematics do not favor such an origin \citep{Mamajek2017, gaidos2018and}. Moreover, the desiccated nucleus that was initially invoked for many of these hypotheses is inconsistent with \om's later-reported non-gravitational acceleration \citep{micheli2018non}. Finally, the theoretical interstellar number densities calculated by \cite{hansen2017postMS} and \cite{moro2018originII} were too low for \om's discovery in PAN-STARRS1 to be statistically favorable.

Here, we reconsider the population of post--main sequence exo-comets in the LSST era. The larger search volume of this impending survey could be amenable to detecting these hypothesized interlopers. If exo-comets that were embedded in the outflows of AGB stars are observationally distinguishable from other planetesimals, then the LSST could constrain the reservoir of these late-ejected interstellar interlopers.

In Section \ref{sec:ejection}, we illustrate the dynamics of ejecting exo-Oort Cloud objects into the Milky Way from post--main sequence stellar mass loss. In Section \ref{sec:processing}, we examine the environmental conditions surrounding AGB stars and the observable signatures that could be imprinted onto small bodies. We develop a model for the occurrence and structure of exo--Oort Clouds in Section \ref{sec:occurrence} and quantify the LSST's sensitivity to these interlopers in Section \ref{sec:LSST}. Combining these results, Section \ref{sec:constraints} elucidates the forthcoming constraints from the LSST on the formation and survival of exo--Oort Clouds. We discuss our results within the context of research on extrasolar planets and already-known interlopers in Section \ref{sec:discussion}. Finally, we summarize our key findings in Section \ref{sec:conclusions}. Following the nomenclature of \cite{hansen2017postMS}, we refer to these hypothesized post--main sequence exo-comets as ``Jurads" for the remainder of this paper.

\section{Ejection of Small Bodies During Post--Main Sequence Evolution} \label{sec:ejection}

Stellar mass loss decreases the radial accelerations of orbiting bodies and shrinks the star's Hill ellipsoid; the combined effect of these processes can unbind exo-comets. \cite{hansen2017postMS, katz2018interstellar} considered the ejection of small bodies from their post--main sequence hosts via close encounters with planets, and \cite{Rafikov2018b} detailed how fragments of tidal disruption events can be ejected from white dwarfs. In contrast, we focus on the pathway explored by \cite{moro2018originII} where the evolving gravitational potential during the AGB phase alone releases minor planets.

Stars must evolve into white dwarfs during the galactic lifetime to generate  Jurads. Assuming (1) that main sequence stars follow a mass-luminosity relationship $L_{\text{MS}} \propto M_{\text{MS}}^{3.5}$ and (2) that the hydrogen budget for core fusion is proportional to the main sequence mass $M_{\text{MS}}$, we approximate the stellar main sequence lifetime by scaling from the solar value:

\begin{equation} \label{eq:tauMS}
    \tau_{\mathrm{MS}} \simeq 10^{10}\,\mathrm{y}\,\,\bigg(\frac{M_{*}}{\mathrm{M}_{\odot}}\bigg)^{-2.5}\,.
\end{equation}

The post--main sequence timescale is short compared to the main sequence, so Equation \ref{eq:tauMS} shows that only $M_{\text{MS}} \gtrsim 0.9\msun$ stars could generate Jurads within $12\,\text{Gyr}$. At our precision, corrections to $\tau_{\text{MS}}$ for stellar metallicity are negligible; stellar mass dictates evolutionary timescales \citep{KWWtextbook}.

\subsection{Dynamical Regime of Exo--Oort Cloud Comets During Post--Main Sequence Evolution}

Without other perturbers, exo-comets evolve as test particles in the variable mass two-body problem during the post--main sequence. The dynamics are parameterized by a dimensionless index $\Psi$ that measures the adaiabaticity of orbital expansion, calculated by comparing the timescales of the stellar mass loss to the orbital period and written as \citep{veras2011escape1}

\begin{equation} \label{eq:adiabaticIndex}
    \Psi \equiv  \, \bigg( \frac{\dot{M}_{*}}{M_{*}} \, \bigg) \bigg(\, \frac{1}{n_{\text{J}}}\, \bigg)\,,
\end{equation}

\noindent where $\dot{M}_{*}$ and $M_{*}$ are the star's instantaneous mass loss rate and mass, respectively, and $n_{\text{J}}$ is the exo-comet's mean motion.

When adiabaticity ($\Psi \ll 1$) applies, the orbit expands along similar ellipses. Thus, the ratio between the initial $a_{\text{J},0}$ and final semimajor axes $a_{\text{J,f}}$ is \citep{veras2011escape1}

\begin{equation}\label{eq:adiabaticSemimajor}
    a_{\text{J,f}} \Big/ a_{\text{J},0} = M_{\text{MS}} \Big/ M_{\text{WD}}\,,
\end{equation}

\noindent where $M_{\text{WD}}$ is the mass of the white dwarf remnant.

With Equation \ref{eq:adiabaticIndex}, we will calculate the adiabaticity of exo--Oort orbits during post--main sequence stellar evolution. We take the inner edge of exo--Oort Clouds to scale from the solar value, with constant orbital period:

\begin{equation} \label{eq:innerOort}
    a_{\mathrm{OC, in}} \simeq 10^{3}\,\mathrm{au}\,\bigg(\frac{M_{\mathrm{MS}}}{M_{\odot}}\bigg)^{1/3}\,.
\end{equation}

Then, we assume that the outer edge follows the host star's Hill sphere radius in the Milky Way for a fixed distance from the galactic center:

\begin{equation} \label{eq:hillSphere}
    a_{\mathrm{H}} \simeq 2.5\times10^{5}\,\mathrm{au}\,\bigg(\frac{M_{*}}{M_{\odot}}\bigg)^{1/3}\,.
\end{equation}

The chronology of $\dot{M_{*}}$ in theoretical AGB evolutionary tracks is often assigned independently of fundamental parameters, but $M_{\text{MS}} \gtrsim 1.3\msun$ stars primarily lose mass during thermal pulses that each transpire over approximately $10^{4}\,\text{yr}$ \citep{KWWtextbook}. Using an empirical relationship \citep{wood1992whiteDwarf},

\begin{equation} \label{eq:WDempirical}
    M_{\mathrm{WD}}\,\simeq \,0.49\,\mathrm{M}_{\odot}\,\,\exp\Bigg({\frac{M_\mathrm{MS}}{10.52\,\mathrm{M}_{\odot}}}\Bigg)\,,
\end{equation}

\noindent the mass lost on the AGB is $\Delta M_{*} = M_{\text{MS}} - M_{\text{WD}}$.

\begin{figure}
    \centering
    \epsscale{1.2}
    \plotone{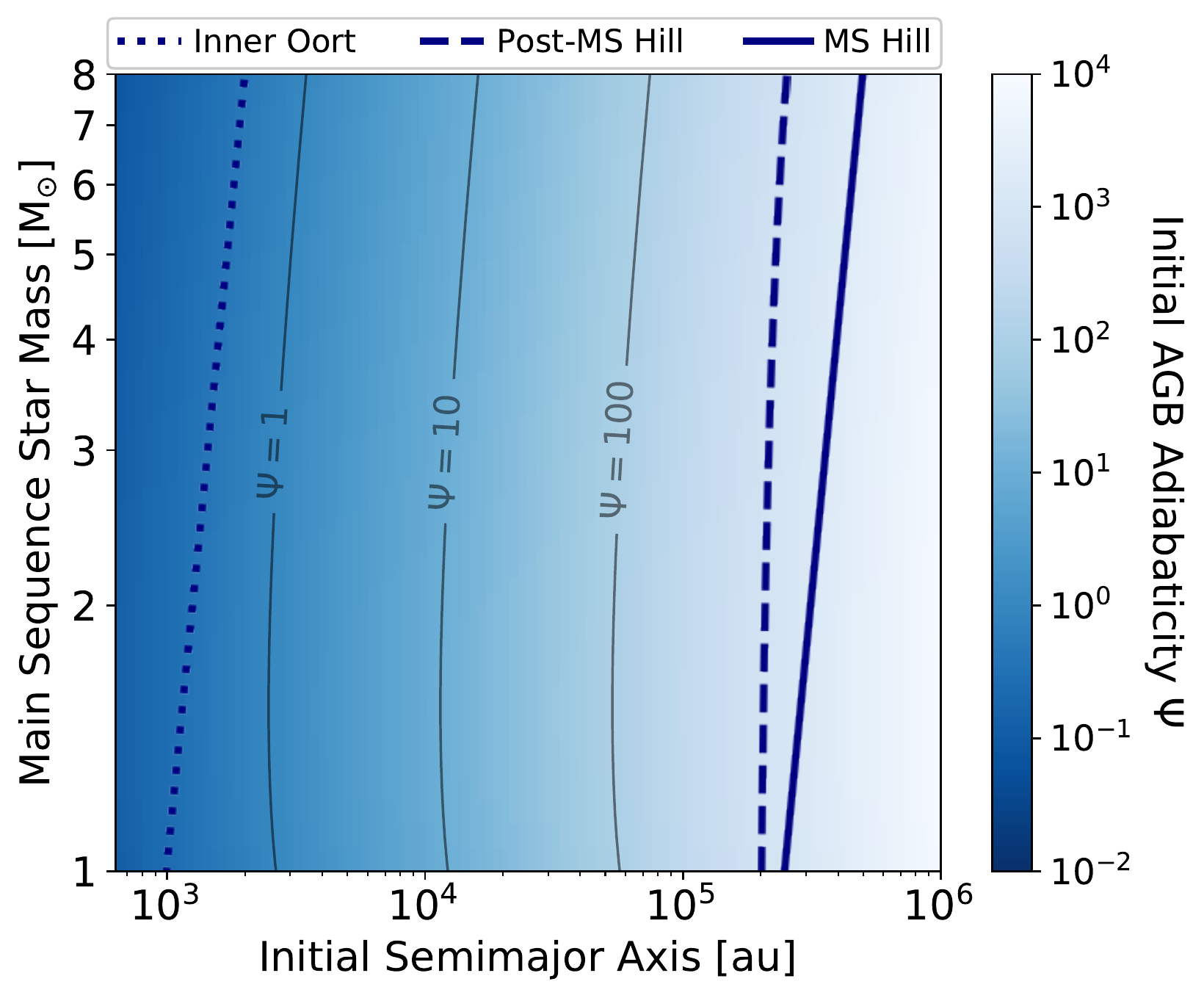}
    \caption{Adiabatic index $\Psi$ (Equation \ref{eq:adiabaticIndex}) for exo--Oort orbital dynamics at the beginning of the host star's AGB. Results are displayed as a heatmap on a grid of main sequence stellar mass versus initial semimajor axis $a_{\text{J}, 0}$. Contours of $\Psi = [1, 10, 100]$ are drawn along with astrocentric distances of interest: the inner exo--Oort Cloud boundary (Equation \ref{eq:innerOort}), the post--main sequence Hill sphere radius, and the main sequence Hill sphere radius (Equation \ref{eq:hillSphere}).}
    \label{fig:oortCloudPsi}
\end{figure}

With constant $\dot{M}_{*}$ during the thermal pulse, we find $\dot{M}_{*} \simeq \Delta M_{*} / (10^{4}\,\text{yr})$ and calculate $\Psi$ at the beginning of the thermal pulse (Figure \ref{fig:oortCloudPsi}). Orbital evolution is non-adiabatic for all but the innermost exo-Oort regions. Since $\Psi$ increases as $M_{*}$ decreases for a constant $\dot{M_{*}}$, Figure \ref{fig:oortCloudPsi} represents the point of maximum adiabaticity during mass loss for our assumed evolutionary track.

\subsection{Analytic Probabilities of Exo-Comet Ejection}

Since most exo--Oort Cloud semimajor axes are firmly in the non-adiabatic regime ($\Psi \gg 1$), the test particle does not complete an orbit during stellar mass loss. We can analytically estimate the fraction of exo-comets that are ejected during the host's AGB phase with the approximation that $\Psi \rightarrow \infty$: instantaneous mass loss where orbital elements change impulsively \citep{hadjidemetriou1966analytic, hadjidemetriou1966binarySystems}. This setup was considered in the context of supernovae in binary star systems \cite{hills1983impulsive}, who derived the following ejection criterion:

\begin{equation} \label{eq:escCriterionSimple}
    \frac{R}{2a_{\text{J}, 0}} < 1 - \frac{M_{\mathrm{WD}}}{M_{\mathrm{MS}}}\,.
\end{equation}

Equation \ref{eq:escCriterionSimple} shows that test particles with a given initial semimajor axis are more likely to become unbound near perihelion. The ejection probability is thus the fraction of the orbital period spent within the the eccentric anomaly range $(2\pi - E_{\text{c}}) < E_{\text{J},0} < E_{\text{c}}$, where $E_{\text{c}}$ is a critical value given by \citep{hills1983impulsive}

\begin{equation} \label{eq:criticalEc}
    E_{\mathrm{c}} = \cos^{-1}\bigg(\frac{2(M_{\mathrm{WD}}/M_{\mathrm{MS}})-1}{e_{\mathrm{J}, 0}}\bigg)\,,
\end{equation}

\noindent where the initial eccentricity is $e_{\text{J},0}$.

Inserting $E_{\text{c}}$ into Kepler's Equation, we find the ejection probability as follows:

\begin{equation} \label{eq:ejectionProbability}
    \mathcal{P}_{\mathrm{ej}} = \pi^{-1}\Big(E_{\mathrm{c}}-e_{\mathrm{J},0}\,\sin({E_{\mathrm{c}}})\Big)\,.
\end{equation}

\begin{figure}
    \centering
    \epsscale{1.2}
    \plotone{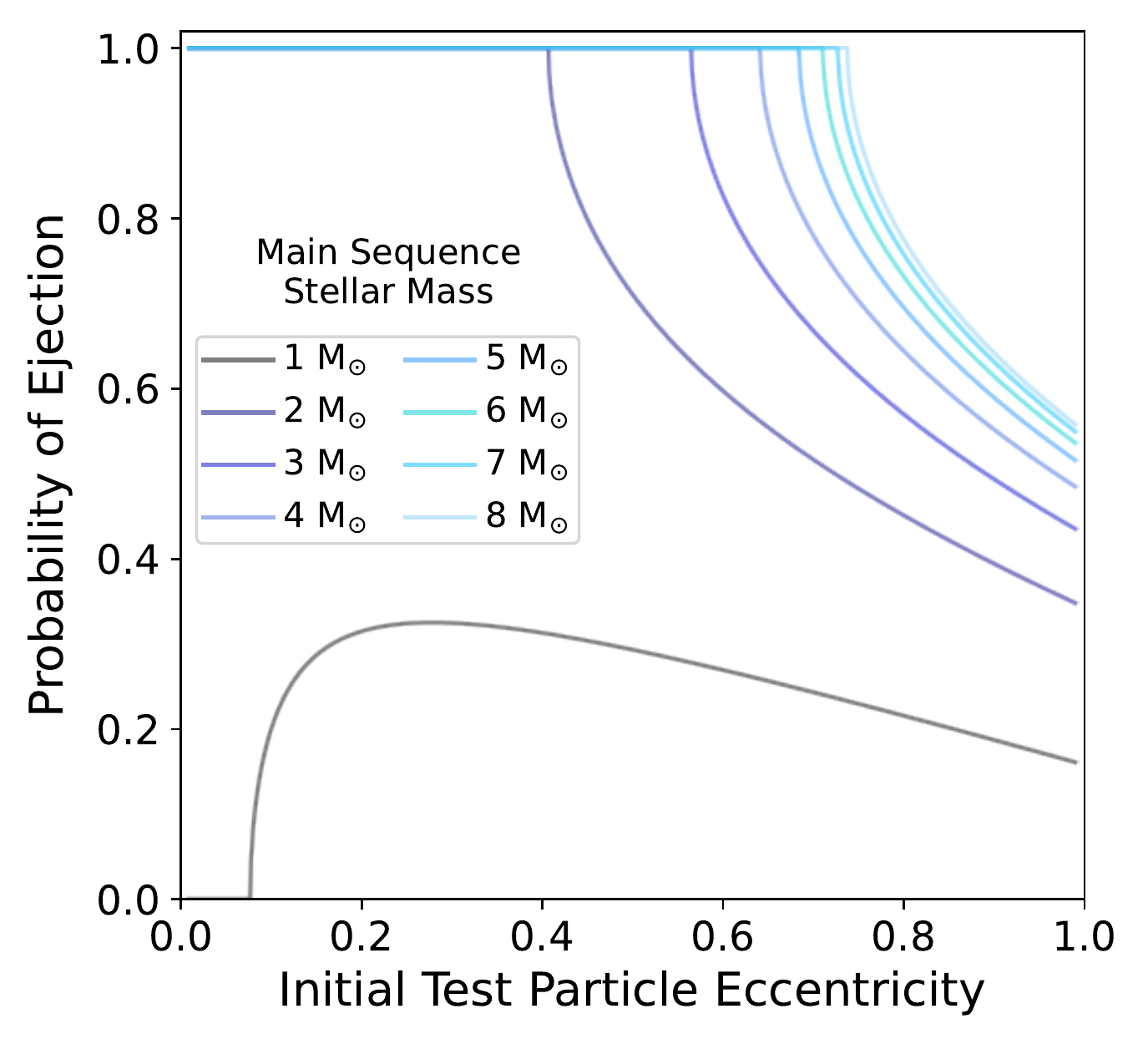}
    \caption{Analytically determined probability (Equation \ref{eq:ejectionProbability}) of ejecting test particles in the non-adiabatic limit where $\Psi \rightarrow \infty$. Equation \ref{eq:WDempirical} determines the total mass lost in the event as $M_{\text{MS}} - M_{\text{WD}}$. Results are plotted against the test particle's eccentricity before the mass loss event, $e_{\text{J},0}$.}
    \label{fig:ejectionProbability}
\end{figure}

Figure \ref{fig:ejectionProbability} shows Equation \ref{eq:ejectionProbability} applied to various $M_{\text{MS}}$, with $M_{\text{WD}}$ calculated with Equation \ref{eq:WDempirical}. In our method, the argument of the inverse cosine in Equation \ref{eq:criticalEc} is manually given a minimum possible value of -1 and maximum possible value of +1 (corresponding to the function's real domain). Since $E_{\text{c}}$ always lies on $[0, \pi]$, the probability in Equation \ref{eq:ejectionProbability} will not exceed unity. Massive stars have larger $\Delta M_{*}$, so the parameter space for which exo-comets are ejected grows with $M_{\text{MS}}$. Stars for which $M_{\text{WD}}/M_{\text{MS}} < 0.5$, corresponding to $M_{\text{MS}} \gtrsim 1.1\,\msun$ according to Equation \ref{eq:WDempirical}, have some initial eccentricities for which test particles are guaranteed to escape.

\subsection{Numerical Simulations of Exo-Comet Escape}

Because Equation \ref{eq:ejectionProbability} applies to $\Psi \rightarrow \infty$, the inner regions of exo--Oort Clouds are not necessarily well-described by the results in Figure \ref{fig:ejectionProbability}. Nonetheless, \cite{veras2011escape1} found that Jurads are ejected from these smaller semimajor axes. To further investigate and validate those findings, we perform N-body simulations in \texttt{rebound} \citep{rebound} for $\Psi \sim 1$. We consider a fiducial $M_{\text{MS}} = 2\msun$ star undergoing a thermal pulse, for which Equation \ref{eq:WDempirical} gives $M_{\text{WD}} = 0.59\,\msun$. Assuming a constant mass loss rate $\dot{M}_{*}$, the stellar mass evolves linearly from time $t \in [0, t_{\text{loss}}]$:

\begin{equation} \label{eq:massLossPrescription}
    M_{*}(t) = M_{\mathrm{MS}} - \bigg(\frac{t}{t_{\mathrm{loss}}}\bigg)\bigg(M_{\mathrm{MS}} - M_{\mathrm{WD}}\bigg)\,.
\end{equation}

\noindent Taking $t_{\text{loss}} = 10^{4}\,\text{yr}$ to be consistent with Figure \ref{fig:oortCloudPsi}, the mass loss rate is $\dot{M}_{*} \simeq 1.4\times 10^{-4} \msun \,\text{yr}^{-1}$. Investigations on the fate of the Sun's Oort Cloud have demonstrated that test particle trajectories are sensitive to the stellar evolution model \citep{veras2012boundary}. Given a constant mass loss fraction $\Delta M_{*}$, however, all mass loss profiles converge to the same ejection fraction for $\Psi \gg 1$ (including the linear one that we adopt).

For the exo-comets, we initialized massless test particles with uniform semimajor axes $a_{\text{J}, 0} \in [500, 4000]\,\text{au}$. Although the smallest values are inside the boundary from Equation \ref{eq:innerOort}, we considered these shorter-period objects to account for uncertainty in the locations of exo--Oort inner edges. Next, we drew these orbital elements from uniform distributions: eccentricities $e_{\text{J},0} \in [0.0, 1.0]$, mean anomalies $l_{\text{J}, 0} \in [0, 2\pi]$, and longitudes of pericenter $\bar{\omega}_{\text{J}, 0} \in [0, 2\pi]$. Spherical symmetry allowed us to set inclinations to zero \citep{veras2011escape1}.

We simulated $10.5\,\text{kyr}$ of evolution with \texttt{rebound}'s IAS15 integrator \citep{reboundias15}, a fifteenth-order and non-symplectic method, although eccentricities and semimajor axes do not change after $t_{\text{loss}} = 10\,\text{kyr}$. Every $t_{\text{s}} = 100\,\text{yr}$ until $t_{\text{loss}}$, we manually changed $M_{*}$ with Equation \ref{eq:massLossPrescription} and resumed the \texttt{rebound} run. We validate the appropriateness of $t_{\text{s}}$ in Appendix \ref{sec:dynamicalModelValidation}. Importantly, the timesteps used internally by \texttt{rebound} were selected by the IAS15 algorithm and were not $t_{\text{s}} = 100\,\text{yr}$.

\begin{figure}
    \centering
    \epsscale{1.15}
    \plotone{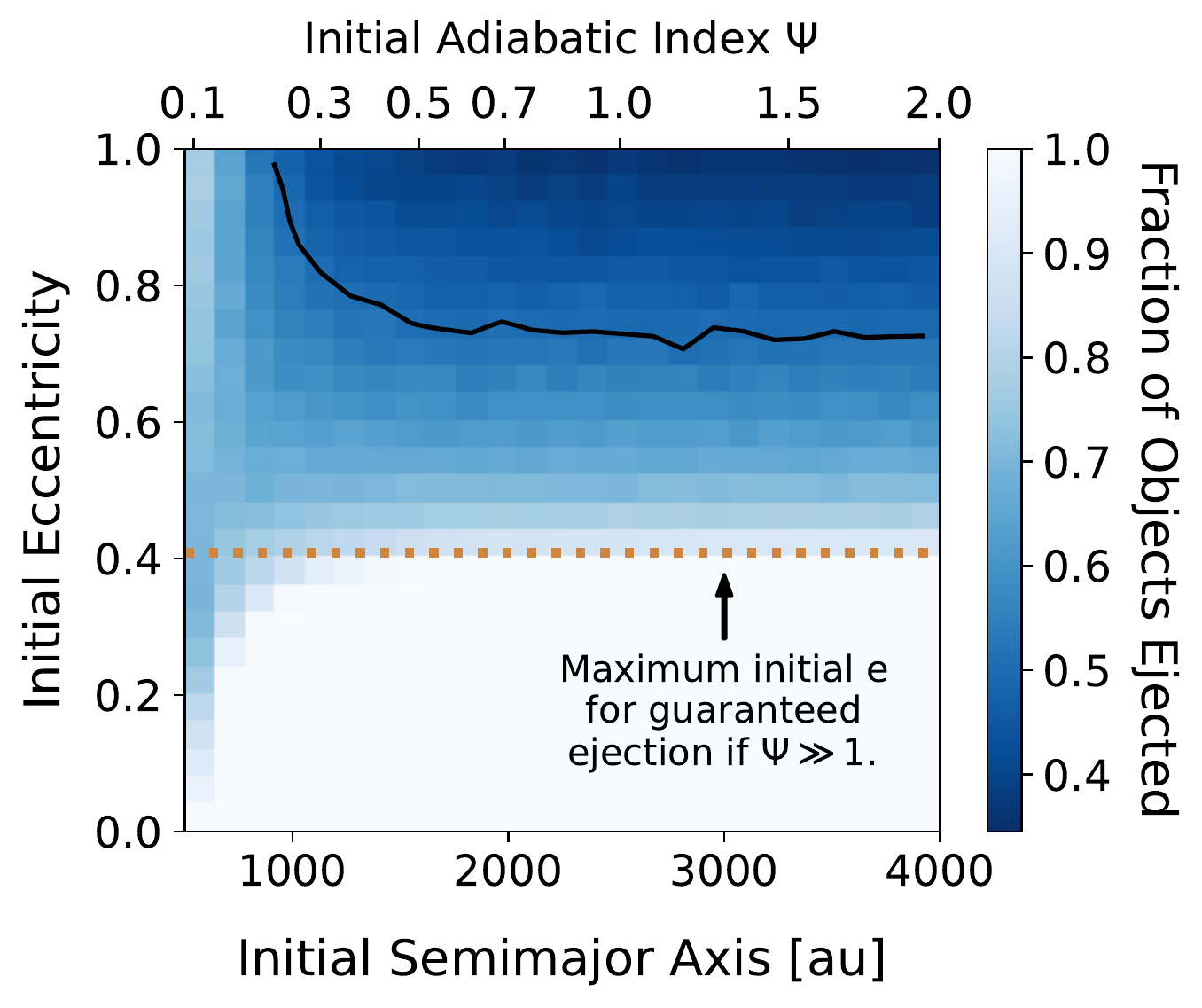}
    \caption{Fraction of test particles that became unbound during \texttt{rebound} simulations of a late-AGB thermal pulse, shown on a 25x25 grid in ($e_{\text{J}, 0}$, $a_{\text{J}, 0}$). Each cell represents approximately 6400 points. The dotted orange line shows the analytic prediction for the $\Psi \gg 1$ regime from Equation \ref{eq:ejectionProbability}: the maximum $e_{\text{J}, 0}$ value for which test particles are guaranteed to become unbound during post--main sequence mass loss. The solid black line shows the contour where half of the test particles become unbound.}
    \label{fig:ejectionStats}
\end{figure}

Having established the veracity of our method, we ran $4\times10^{6}$ test particles with the aforementioned initial distributions of orbital elements. We find that 74\% of the exo-Oort Cloud test particles are ejected, which corresponds nicely with the eccentricity-averaged fraction of 73\% for the same fiducial $M_{\text{MS}} = 2\,\msun$ case from the analytic results on Figure \ref{fig:ejectionProbability}. We display the numerical ejection fraction as a function of $a_{\text{J},0}$ and $e_{\text{J},0}$ in Figure \ref{fig:ejectionStats}, confirming that higher initial eccentricity leads to lower ejection fractions. The eccentricity-averaged ejection fraction is nearly constant across all semimajor axes that we consider, consistent with the finding by \cite{veras2011escape1} that $\Psi \gtrsim 0.02$ spurs non-adiabatic dynamics. On Figure \ref{fig:ejectionStats}, we show that our numerical results converge to the limiting case from Equation \ref{eq:ejectionProbability}. In total, these \texttt{rebound} simulations show that small bodies from even the inner regions of exo--Oort Clouds (where $\Psi \sim 1$) are ejected during the post--main sequence.

\subsection{Other Dynamical Considerations}

Given the consistent results from Figures \ref{fig:ejectionProbability} and \ref{fig:ejectionStats}, we assume an overall ejection fraction of $f_{\text{ej}} = 0.7$ for exo-Oort comets orbiting $1-8\,\msun$ thermally-pulsing AGB stars. Our simplifying liberties often underestimate Jurad ejection. For example, Figure \ref{fig:ejectionStats} assumed $M_{\text{MS}} = 2\msun$ even though more massive stars generate more Jurads per Figure \ref{fig:ejectionProbability}. We also assume isotropic mass loss despite observations of AGB outflows \citep{sahai2003collimatedAGBoutflow} revealing collimated and aspherical morphologies. Work by others \citep{hills1983impulsive, parriott1998anisotropic} found that these asymmetries eject more exo-comets in the non-adiabatic regime.

Thus far, we have considered single stars. Detailed studies by \cite{veras2012escape2} and \cite{veras2014escape3} found that exo-comets are more likely to escape from post--main sequence binaries than from single stars, especially for wide-separation and eccentric stellar companions. Notably, binary stars have been proposed as prolific producers of interstellar small bodies \citep{Cuk2017, Jackson2017, childs2022misalignedDiskISOs} during the (pre--)main sequence. Planetary perturbers could similarly affect ejection fractions, and indeed a population of giant planets is required to generate the initial exo-Oort Clouds \citep{dones2004oort}. We discuss the relationship between Jurads and exoplanets in Section \ref{sec:discussion}.

Galactic tides and stellar flybys can eject exo--Oort Cloud comets anytime during stellar evolution \citep{veras2011escape1, moro2018originII}. Small bodies lost before the post--main sequence contribute to the Milky Way's rogue minor planets but are not considered ``Jurads," per our definition. For the Sun, \cite{heisler1986galacticTide} found that the galactic tide strips more comets than do stellar flybys and that the survival lifetimes are of order $1\,\text{Gyr}$ against ejection \citep{hut1985oortCloud, dones2004oort}. While a majority of the Sun's original outer Oort Cloud may destabilize during the main sequence, the intermediate mass stars which we consider have comparatively brief lifetimes. This shortened time-integrated erosion could imply that a larger fraction of $2-8\,\msun$ star exo--Oort Clouds are intact at the end of the main sequence, although detailed modeling is required to vet this possibility.

Taking these other considerations into account, we believe that $f_{\text{ej}} = 0.7$ is a fair estimate of the fraction of exo-Oort Cloud objects which are retained throughout the star's main sequence lifetime and are subsequently ejected during the post--main sequence. This value is similar to a previous estimate of $40\%$ \citep{moro2018originII}, determined by extrapolating from results published by \cite{veras2014escape3}. At the population-level, post--main sequence systems will generate Jurads from any exo-Oort Cloud objects that do exist. Should $f_{\text{ej}}$ be different, the detection prospects for the LSST (Section \ref{sec:LSST}) would change linearly.

\section{Exo-Comet Processing During Late-Stage Stellar Evolution} \label{sec:processing}

For Jurads to serve as tracers of exo--Oort Cloud occurrence, these small bodies must be distinguishable from their counterparts that were ejected before the post--main sequence. Because rogue small bodies are expected to mimic the velocity dispersions of their progenitor stellar populations \citep{Jewitt2022}, the inbound velocities of Jurads towards the solar system should resemble those of local white dwarfs. Other interstellar interlopers, in contrast, should have kinematics drawn from younger stellar populations. Because these distributions overlap, however, kinematics alone will not confirm the Jurad nature of an exo-comet. Therefore, we explore the possibility that minor planets in post--main sequence environments might bear observational signatures of their history.

The thermal evolution of (exo-)comets was examined by \cite{stern1988flyby} and \cite{stern1990PMS}, who considered heating of the Sun's Oort Cloud via stellar flybys and the heating of Kuiper Belt Objects from luminous post--main sequence stars, respectively. The latter study \citep{stern1990PMS} calculated destruction timescales for water ice and aggregate outgassing rates from comet clouds, with the idea of possibly explaining the presence of H$_2$O masers and complex molecules in AGB environments.

In the LSST era, our study advances this research to consider the processing imparted to individual exo-comets that may be observed as interlopers passing through our solar system. The solar Oort Cloud is believed to be our system's largest reservoir of cometary material, so we focus on these semimajor axes for extrasolar minor planets. Motivated by Borisov's hypervolatile-rich coma, we consider the viability of these tenuous substances surviving during the host star's post--main sequence. During stellar flybys, \cite{stern1988flyby} found that heat would only penetrate the top $10\,\text{m}$ of exo-comets due to low thermal diffusivity. The post--main sequence lasts more than a factor of $10^{4}$ longer than stellar flybys; in the present context, we will show that hypervolatiles can be heated above nominal sublimation points throughout km-scale Jurads.

\subsection{Processing from Stellar Luminosity}\label{subsec:thermprocess}

Stellar luminosities increase by orders-of-magnitude during post--main sequence evolution, and the effective temperature at distance $R$ from the host star is

\begin{equation} \label{eq:effectiveTemperature}
    \teff = \bigg(\frac{L_{*}(1-\mathcal{A})}{16\pi\sigSB R^{2}}\bigg)^{1/4}\,,
\end{equation}

\noindent where $\mathcal{A}$ is the Bond albedo, $L_{*}$ is the stellar luminosity, and $\sigSB$ is the Stefan-Boltzmann constant. 

In the solar Oort Cloud, $\teff \simeq 3\,\text{K}$ during the main sequence. The energy balance has comparable components from the cosmic microwave background and solar flux \citep{umurhan2022arrokoth}. During the AGB, however, $L_{*}$ will increase to several $10^{3}\,\lsun$. Because the star does not inject a significant amount of dust into the environment during the early-AGB phase, we assume that the interplanetary medium remains optically thin until the thermal pulses begin. Therefore, Equation \ref{eq:effectiveTemperature} is valid for $\teff$ in exo--Oort Clouds through the early-AGB.

During the Sun's AGB phase, $\teff \gtrsim 40\,\text{K}$ at $4000\,\text{au}$ for $L_{*} = 5000\,\text{L}_{\odot}$. At the order-of-magnitude level, the depth $d$ to which this heat penetrates is $d \sim (t_{\text{c}}\alpha_{\text{t}})^{1/2}$, where $\alpha_{\rm t}$ is the thermal diffusivity of the material \citep{stern1988flyby}. Common dielectric solids such as rock and ice have $\alpha_{\text{t}} \sim 10^{-6}\,\text{m}^{2}\,\text{s}^{-1}$, but porous materials have $\alpha_{\text{t}} \sim 10^{-8}\,\text{m}^{2}\,\text{s}^{-1}$ \citep{Jewitt2017}. Setting $\alpha_{\rm t}=10^{-8}$ m$^2$ s$^{-1}$ and $t_c=10^7\,\text{yr}$, we find $d \sim 2\,\text{km}$ during the AGB phase. For the icy or rocky case with $\alpha_{\rm t}=10^{-6}\,{\rm m}^2\,{\rm s}^{-1}$, we find $d \sim 20\,{\rm km}$. Even with a lower bound on the thermal diffusivity represented by the porous case, the temperature should converge to $\T_{\rm eff}$ at all depths for minor planets up to an order-of-magnitude larger than \om.

To confirm this assessment that inner exo--Oort Cloud Jurads will lose hypervolatiles, we perform numerical simulations tracking the thermal evolution of Jurads during the post--main sequence. We assume a spherically symmetric exo--Oort Cloud and an isotropic radiation field; the stellar evolution timescales are much longer than any observed rotation period for small bodies. By considering just the effective temperature, we ignore possible effects of obliquity on the volatile evolution of small bodies \citep{schorghofer2008mainBeltIce}. Then, we solve the 1-D heat equation for a spherical small body,

\begin{equation}\label{eq:heat}
    \frac{\partial \T}{\partial t}=\bigg(\frac{\kappa}{\rho c_P}\bigg) \frac{1}{r^2}\frac{\partial}{\partial r}\bigg(r^2\frac{\partial \T}{\partial r}\bigg)\,.
\end{equation}

\noindent Equation \ref{eq:heat} is a partial differential equation for temperature $\T$ that depends on radius $r$ and time $t$, where $\kappa$ is the thermal conductivity, $c_P$ is the specific heat capacity, and $\rho$ is the bulk density. 

\begin{figure}
    \centering
    \epsscale{1.2}
    \plotone{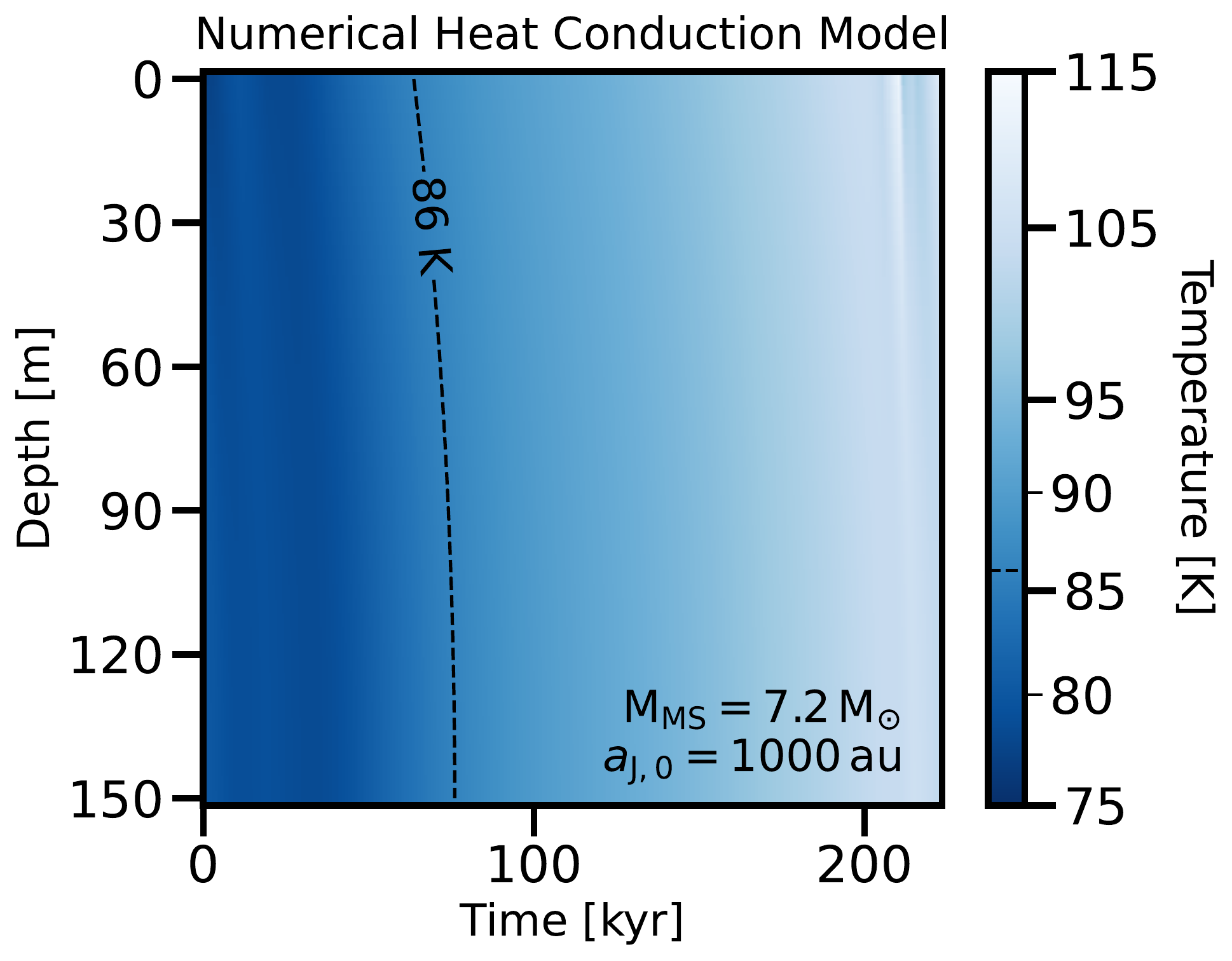}
    \caption{Output of a numerical thermal penetration simulation for an object orbiting a $7.2\msun$-mass star at 1000 au. The temperature is initialized as $\teff$ at the end of the RGB and is evolved through the early- and thermally pulsing--AGB. The temperature spike} in the upper right-hand side corner is the increase in stellar luminosity during the first thermal pulse. Of all stellar masses under consideration, this one precipitates the steepest temperature gradient across the minor planet. With only $\Delta \T\sim 10$K from the surface to the center, our use of effective temperature and isothermality for the exo-comet's temperature is justified.
    \label{fig:AGBheatSim}
\end{figure}

The numerical method of lines (MOL) \citep{nummethods} is used to solve partial differential equations such as Equation \ref{eq:heat}. By discretizing the spatial dimension, we transform Equation \ref{eq:heat} into a set of ordinary differential equations (ODEs) which can be solved by any ODE method --- for our case, we used the 4th-order Runge-Kutta (RK4) \citep{nummethods}. The right-hand side of Equation \ref{eq:heat} at a given time step is computed using the central finite difference methods for both first and second order derivatives. The spatial derivative of $\T$ at a spatial index $i$ is given by

\begin{equation}\label{eq:finitediff}
    \frac{d \T_i}{dt} = \frac{\alpha}{\Delta r^2}\bigg[ (1+\frac{\Delta r}{r_i})\T_{i+1} + (1-\frac{\Delta r}{r_i})\T_{i-1}-2\T_i\bigg]\,,
\end{equation}

\noindent where $\alpha=\kappa/(c_P \rho)$ is the thermal diffusivity.

The two fictitious points (alternatively called ``ghost zones") in this set of equations, one at each boundary ($\T_{-1}$ and $\T_{N+1}$), are resolved via the Neumann boundary conditions --- symmetry at the center, and radiation heat flux at the surface. At the center we set $\T_{-1}=\T_1$. At the surface, we impose the heat flux from stellar radiation field. Therefore, our boundary conditions are
 
\begin{equation}
\frac{\partial \T}{\partial r}=
\begin{cases}
    \frac{1}{\kappa}\bigg[\frac{\zeta(1-\mathcal{A}) L_*}{4\pi a^2}-\epsilon\sigSB \T^4\bigg] & {\rm at }\, r=R \\
    0 & {\rm at }\, r=0
\end{cases}\,,
\end{equation}

\noindent where $\zeta=0.25$ is the effective projected surface area, and $\epsilon$ is the emmissivity. This equation relates the fictitious points to domain points, closing our set of ODEs and allowing us to apply RK4. 

We performed extensive testing and found a stringent Courant-Friedrichs-Lewy (CFL) criterion of $\alpha \Delta t/\Delta r^2 < 0.7$. With this numerical consideration in mind, we calculate the thermal evolution of exo-comets. We assume $\rho=0.5$ g cm$^{-3}$ and $c_P=2000$ J kg$^{-1}$ K$^{-1}$, typical values for cometary ices \citep{britt2006cometprops}. We use $\kappa=10^{-2}$ W K$^{-1}$ m$^{-1}$, conservatively taking the lower bound between  \cite{gundlachblum} and \cite{steckloff_arrokoth}. Stellar luminosity models are taken from the MESA Isochrones and Stellar Tracks (\href{https://waps.cfa.harvard.edu/MIST/model\_grids.html}{MIST}) database \citep{MIST0, MIST1, MESA1, MESA2, MESA3}.

The exo-comet's thermal conditions at the beginning of the AGB phase are set with $\T_{i} = \teff$ from the end of the red giant branch (RGB) for all $i$. We validate this isothermality by numerically evolving a $r = 150\,\text{m}$ exo-comet placed at $1000\,\text{au}$ from a $2\msun$ RGB star. During the approximately $10^{7}\,\text{yr}$ RGB phase, the minor planet assumes $\teff$ at all depths. These results confirm our order-of-magnitude scaling relationship, $t_{c} \sim d^2/\alpha$, where $T(r) = \teff$ after $t_{c} \sim10^{5}$ years. 

\begin{figure}
    \centering
    \epsscale{1.2}
    \plotone{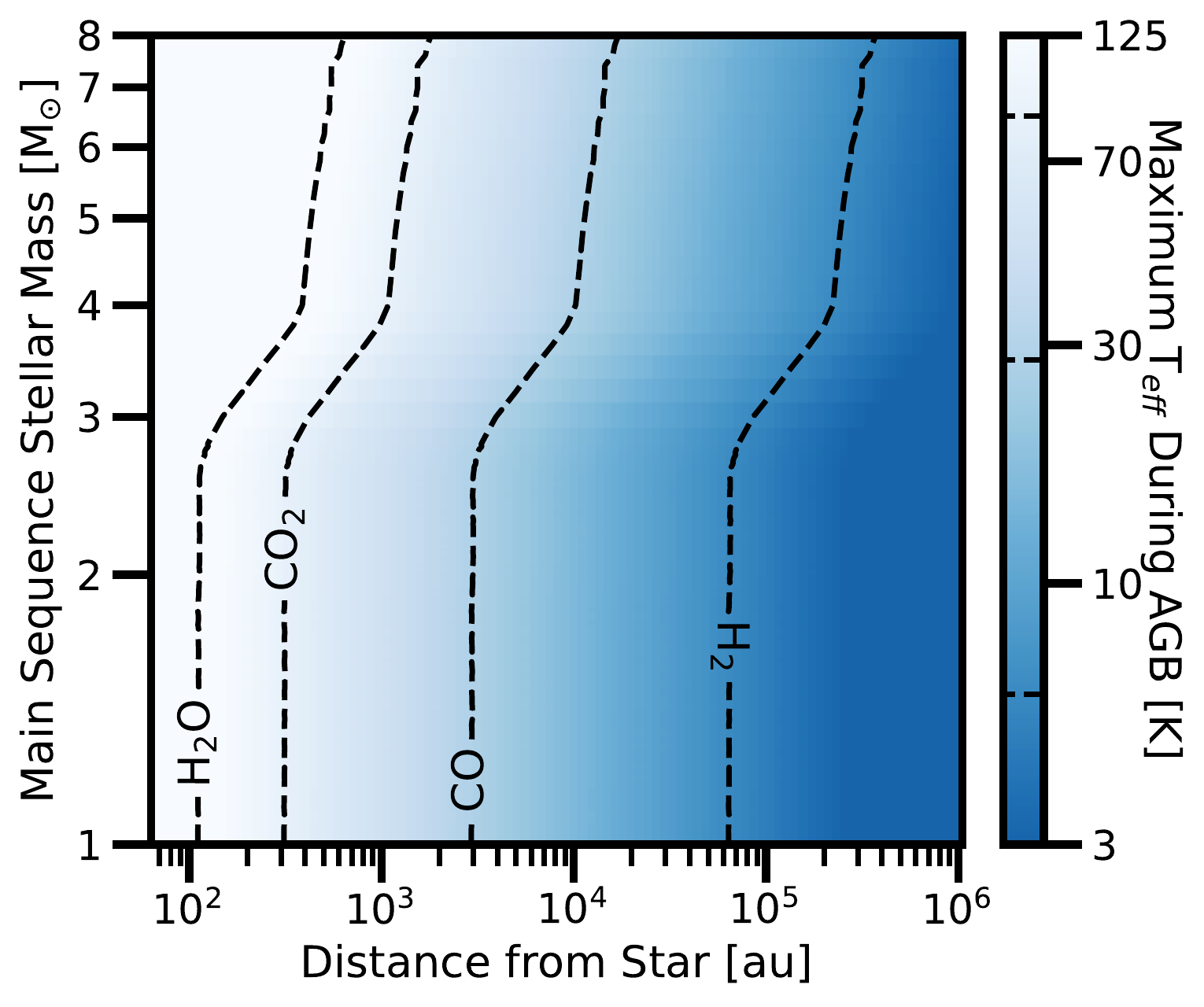}
    \caption{Maximum $\teff$ during the AGB phase as a function of stellar mass and astrocentric distance. Contours at $6\,\text{K}$, $28\,\text{K}$, $86\,\text{K}$, and $144\,\text{K}$ indicate representative values for which H$_2$, CO, CO$_2$, and H$_2$O can be destroyed, respectively. All contour levels come from \cite{Gasc2017} except for H$_{2}$, which comes from \cite{Seligman2020}.}
    \label{fig:oortCloudTemperatures}
\end{figure}

Using these initial conditions, we evolve Equation \ref{eq:heat} for a range of stellar masses and exo-comet semimajor axes ($1-8\,\msun$ and $1000-128000 \,\text{au}$) through the AGB stages. The simulations are halted when the star has lost 10\% of $M_{\text{MS}}$. At that point, the aeolian effects from the stellar mass outflows will complicate the thermal modeling. Figure \ref{fig:AGBheatSim} shows the time-dependent thermal profile of a Jurad around a $7.2\,\msun$ star with $R = 1000\,\text{au}$. Despite being among the most rapidly-evolving stars in our range of considered masses, the temperature gradient across the $150\,\text{m}$ exo-comet remains small throughout the simulation, with $\Delta \T \simeq 10\,\text{K}$. We verified that other stellar mass tracks induce comparable or shallower temperature gradients (not shown), demonstrating that the exo-comet is isothermal at $\teff$ when exposed to the radiation field. Figure \ref{fig:oortCloudTemperatures} shows the maximum $\teff$ as a function of astrocentric distance $R$ and $M_{\text{MS}}$, which we assume are the temperature of the Jurad at all depths.

Contours on Figure \ref{fig:oortCloudTemperatures} delineate representative sublimation temperatures for interstellar ices. These temperatures represent the unrealistic scenario of pure compositions; real cometary ices will be mixed. The contour shapes are determined by the dependence of the maximum stellar temperature on the stellar mass. For $M_\star<3\msun$, there is only a weak dependence. In contrast, the maximum temperature increases sharply between 3-4$\msun$ and continues to increase for $M_\star>4\msun$. Sublimation rates depend exponentially on the ices' binding energies and these species' access to the surface of the small body. Exo-comet volatiles that would nominally be lost to interstellar space in the pure form can be retained under astrophysically-relevant timescales \citep{bergner2023H2}. For example, main belt comets may preserve ice beneath a ``buried snow line" for a range of approximately $10\,\text{K}$ near a nominal sublimation point depending on the properties of the surface dust \citep{schorghofer2008mainBeltIce}. Nonetheless, Figure \ref{fig:oortCloudTemperatures} shows that inner exo--Oort comets may become depleted in hypervolatiles like CO during their host stars' AGB phases. Like \cite{stern1990PMS}, we find that water ice can only sublimate within $R \lesssim 100\,\text{au}$.

\subsection{Processing from Stellar Winds}

Besides stellar luminosity, another possible source of exo-comet processing in post--main sequence environments is the stellar outflows from thermally-pulsing AGB stars. During the main sequence, the stellar wind generates a bubble in the interstellar medium: the ``astrosphere," in analogy to the Sun's heliosphere. The Voyager probes crossed the heliopause about $120\,\text{au}$ from the Sun \citep{gurnett2019voyager}, but outflows from late AGB stars expand this cavity into the exo--Oort region \citep{draine2011book}. Here, we assess the persistence of exo-comet volatiles through the shock itself and ablation from the ensuing stellar wind.

Assuming free expansion into a pressureless surrounding, we can write the maximum astrocentric distance reached by the wind as \citep{draine2011book}

\begin{equation} \label{eq:freeExpansionRadius}
    R_{\mathrm{exp}} \simeq 1.7\times 10^{6}\,\mathrm{au} \bigg(\frac{\dot{M}_{*}}{10^{-4}\,\msun\,\mathrm{yr}^{-1}}\bigg)^{1/2}\bigg(\frac{n_{\mathrm{g}}}{1\,\mathrm{cm}^{-3}}\bigg)^{-1/2}\,,
\end{equation}

\noindent where $n_{\text{g}}$ is the number density of gas particles that are upstream of the wind.

The expansion timescale is \citep{draine2011book}

\begin{equation} \label{eq:freeExpansionTime}
    \begin{split}
    t_{\mathrm{exp}} \simeq 1.5\times10^{5}\,\mathrm{yr} \bigg(\frac{\dot{M}_{*}}{10^{-4}\,\msun\,\mathrm{yr}^{-1}}\bigg)\\\bigg(\frac{v_{\mathrm{w}}}{30\,\mathrm{km}\,\mathrm{s}^{-1}}\bigg)^{-3/2}\bigg(\frac{n_{\mathrm{g}}}{1\,\mathrm{cm}^{-3}}\bigg)^{-1/2}\,,
    \end{split}
\end{equation}

\noindent where $v_{\text{w}}$ is the shell's expansion velocity. In Equations \ref{eq:freeExpansionRadius} \& \ref{eq:freeExpansionTime}, we have assumed fiducial conditions for late AGB stars. The shock front moves through the entire Hill sphere and reaches all exo-Oort comets. For the aforementioned outflows, the Mach number is $\mathcal{M} \simeq 10$.

Hypervolatiles on long-period comets are observed to survive through the heliopause \citep{biver2018c2016r2, mckay2019c2016R2}, leading to the prediction that the astropauses of AGB stars will leave Jurads similarly unaffected. We can corroborate this hypothesis via an order-of-magnitude estimate of the timescale over which a given exo--Oort Cloud object is subject to the shock.

Dust and molecular species are entrained in AGB outflows, and these components cool the gas within a few collisional timescales. The mean free path is

\begin{equation} \label{eq:mfp}
    l_{\mathrm{mfp}} \sim (n_{\mathrm{g}}\sigma_{\mathrm{g}})^{-1}\,.
\end{equation}

In Equation \ref{eq:mfp}, $\sigma_{\text{g}} \simeq 10^{-15}\,\text{cm}^{2}$ is the collisional cross-section of the gas particles. This length scale corresponds to the thickness of the shock transition zone and is of order $60\,\text{au}$ for the outflows that we consider. The shock front passes the exo-comet in approximately $10\,\text{yr}$.

Kinetic energy transfer from collisions between gas particles and the exo-comets is the mechanism by which energy is imparted to Jurads from the outflow in both the shock transition zone and the stellar wind, with the bulk of the kinetic energy transported by ions \citep{Noguchi2011}. Continuing to assume isotropic mass loss, the outflow per cross-sectional area at a given stellar distance is $\Delta M_{*}/(4\pi R^{2})$. Therefore, the total mass that strikes a spherical exo-comet ($M_{\rm gas}$) is

\begin{equation}
    M_{\rm gas} \simeq \Delta M_{*}\,\bigg(\frac{r^2}{4R^2}\,\bigg)\,,
\end{equation}

\noindent where $r$ is the small body's radius and $R$ is the distance from the host star, consistent with Section \ref{sec:ejection}. The total kinetic energy delivered by the stellar wind to an exo-comet, $E_{\rm wind}$, can be written as

\begin{equation} \label{eq:windEnergy}
    E_{\rm wind}\simeq\frac{\Delta M_{*} v_{\rm w}^2}{8}\frac{r^2}{R^2}\,.
\end{equation}

\noindent In Equation \ref{eq:windEnergy}, we have assumed the complete transfer of kinetic energy from the outflow to the comet to find an upper bound on volatile destruction.

We consider small bodies with volatile species of average atomic mass $X$, ice density $\rho_{\rm ice}$, and number-averaged enthalpy of sublimation $\Delta H$. The total energy required to sublimate all of the volatiles contained within the exo-comet, $E_{\rm ice}$, is

\begin{equation} \label{eq:iceEnergy}
     E_{\rm ice}\simeq \frac{4\Delta H \rho_{\rm ice}\pi R^3}{3X}\,.
\end{equation}

\begin{figure*}
    \centering
    \plotone{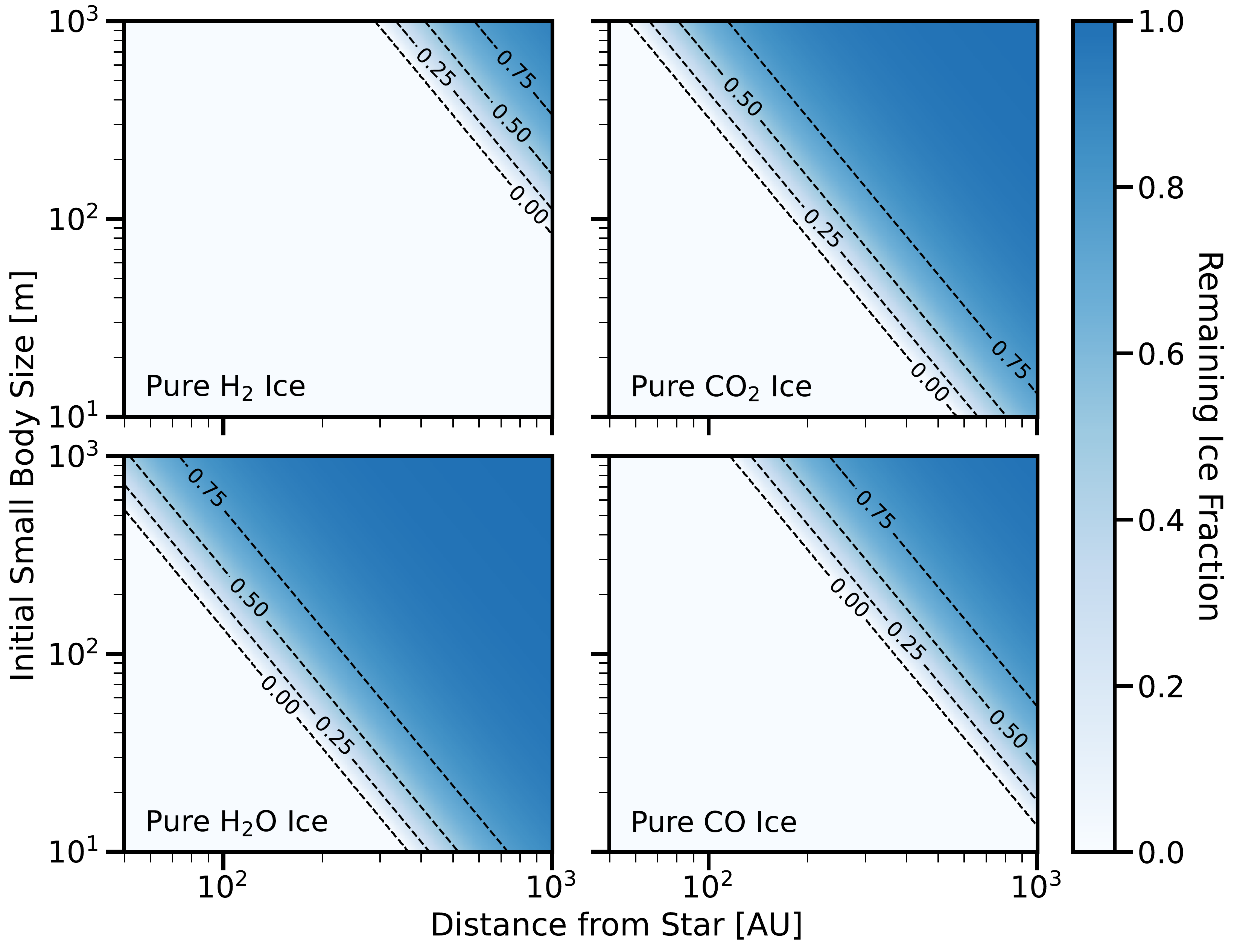}
    \caption{Ice retention fractions against the AGB wind for small bodies composed of pure substances. The contours (from left-to-right) represent surviving ice fractions $f \in [0.0, 0.25, 0.5, 0.75]$. Importantly, the exo--Oort Clouds under consideration in this study have an inner edge near $1000\text{au}$ (the upper limit of this figure).}
    \label{fig:remainingIceFraction}
\end{figure*}

For an order-of-magnitude estimate, we find the fraction of ice destroyed by the wind from comparing the energy scales in Equations \ref{eq:windEnergy} and \ref{eq:iceEnergy}, defining the remaining mass fraction as $f_{\rm r} = 1-E_{\rm wind}/E_{\rm ice}$. With this framework, we calculate $f_{\text{r}}$ for idealized, single-volatile small bodies (Figure \ref{fig:remainingIceFraction}). We use $v_{\rm w}\simeq 30$ km s$^{-1}$ and the fiducial $2 \msun$ star from Section \ref{sec:ejection}, where $\Delta M_{*} \simeq 1.41\msun$. Our input chemical properties are provided in Table \ref{tab:volatileproperties}. In this calculation, we ignore the thermal effects described in the previous subsection and consider the stellar wind in isolation.

For all volatile species, negligible ice is destroyed beyond $1000\,\text{au}$ since the stellar wind mass flux scales with $R^{-2}$. Figure \ref{fig:remainingIceFraction} shows the astrocentric distance ranges where $f_{\rm r}$ transitions from 1 to 0, as the behavior is consistent beyond the visible range. While the decameter-scale H$_2$ and CO objects at the inner edge of the Oort cloud would be destroyed, the majority of the relevant parameter space survives with a full volatile complement. The kinetic energy transferred from AGB winds to exo-comets does not meaningfully process Jurads; thermal effects are more important.

\begin{table}[]
\centering
\caption{Volatile properties from \citet{Seligman2020} (H$_2$, H$_2$O, CO$_2$), \href{https://webbook.nist.gov/cgi/inchi/InChI\%3D1S/CO/c1-2}{NIST} (CO) and \citet{Luna2022} (CO).}
\begin{tabular}{|c|c|c|c|c|}
\hline
\textbf{} & \textbf{$\rho$} & \textbf{$\Delta H$} & \textbf{X} & \textbf{T$_{\text{sub}}$} \\ \hline
Units & [$\text{g}\,\text{cm}^{-3}$] & [$\text{kJ}\,\text{mol}^{-1}$] & [$\text{g}\,\text{mol}^{-1}$] & [K] \\ \hline
H$_{2}$ & 0.08 & 1 & 2.016 & 6 \\ \hline
H$_{2}$O & 0.82 & 54.46 & 18 & 155.0 \\ \hline
CO$_{2}$ & 1.56 & 28.84 & 44 & 82 \\ \hline
CO & 0.85 & 8.1 & 28 & 60 \\ \hline
\end{tabular}
\label{tab:volatileproperties}
\end{table}

Although kinetic processing from the stellar wind does not impart observable signatures onto Jurads, outflowing material could be deposited onto the small bodies themselves. AGB envelopes exhibit astrophysically-unique chemical signatures \citep{ziurys2006AGBchemistry}, so the characteristic polycyclic aromatic hydrocarbons (PAHs) and other carbon-enriched compounds could be diagnostic of an interstellar interloper's Jurad origin if detectable. We can calculate the fraction of the exo-comets' surfaces that would be covered under the following assumptions:

\begin{enumerate}
    \item The AGB outflow is isotropic with metallicity $Z_{\text{w}}$.
    \item Metals in AGB outflows that do impact Jurads stick with an efficiency $f_{\text{st}}$ and are uniformly distributed across the surface.
    \item Dust grains have uniform radii $r_{\text{d}}$, which is also their typical spacing on the exo-comet surface. The bulk density of AGB grains is $\rho_{\text{d}}$.
\end{enumerate}

With this setup, we find the fraction of an exo-comet's surfaces that are covered by AGB dust as

\begin{equation} \label{eq:dustCoverage}
    f_{\text{d}} = \frac{3Z_{\mathrm{w}} \Delta M_{*} f_{\mathrm{st}}}{64\pi r_{\mathrm{d}}R^{2}\rho_{\mathrm{d}}}\,.
\end{equation}

Using fiducial values $Z_{\text{w}} = 0.02$, $r_{\text{d}} = 1\,\mu\text{m}$, $\rho_{\text{d}} = 1\,\text{g}\,\text{cm}^{-3}$, $R = 1000\,\text{au}$, and $f_{\text{st}} = 0.5$, we find that Jurads from the inner Oort Cloud could be covered by tens of layers of dust originating from the host AGB star if solids in the outflow are efficiently accreted.

\section{Occurrence of Exo--Oort Comets} \label{sec:occurrence}

To evaluate the prospects for identifying a Jurad in the LSST, we first estimate the occurrence of such objects. We appraise the reservoir of solids in exo--Oort Clouds and divide this material into small body populations with a range of assumed (and currently unconstrained) size-frequency distributions (SFDs). Then, we model the expected occurrence of Jurads by extrapolating from the population-level kinematics of local white dwarfs.

\subsection{Evaluating the Solid Mass Per Star}

In the absence of observational constraints on exo--Oort Cloud formation, mass, and SFDs, we construct an \textit{ab initio} model towards assessing the likelihood of detecting Jurads with the LSST. We assume that the initial reservoir of minor planets  scales with the host star's metallicity, the available planetesimal-building material in the primordial environment and that the solar system's fractional partitioning of minor bodies into the Oort Cloud and other reservoirs (including interstellar space) is representative of the Jurad progenitor systems. Taking a constant mass ratio between circumstellar disks and host stars, the exo--Oort mass reservoir should scale linearly with stellar mass and metallicity. Although this assumption may not hold on a system-by-system basis, we are interested in a galactic average to inform the LSST detectability of Jurads.

To determine the mass of a given star's Jurad ejecta, we also require the fraction of initially-formed exo--Oort Cloud objects that are retained through the star's main sequence lifetime and subsequently ejected during the AGB stage: $f_{\text{ej}} \simeq 0.7$ from Section \ref{sec:ejection}. From these components, we formulate our estimate for the mean mass of Jurad ejecta from a typical star as

\begin{equation} \label{eq:exoOortMass}
    \begin{split}
    M_{\mathrm{J}, *} = 0.56{\rm M}_{\oplus}\bigg(\frac{2\,\mathrm{M}_{\oplus}}{\mathrm{M}_{\mathrm{OC}, \odot}}\bigg)\bigg(\frac{M_{\mathrm{MS}}}{M_{\odot}}\bigg)\\ \bigg(\frac{10^{[\mathrm{Fe/H}]}}{0.4}\bigg)\bigg(\frac{f_{\mathrm{ej}}}{0.7}\bigg)\,.
    \end{split}
\end{equation}

\noindent In Equation \ref{eq:exoOortMass}, $[\text{Fe/H]}$ is the stellar metallicity relative to the Solar value in dex and $M_{\text{OC},\odot}$ is the assumed mass of the solar system's Oort cloud. On average, $\langle [\text{Fe/H}] \rangle = -0.4$ for the thin disk that holds most of the galaxy's stars \citep{freeman2002reviewGalaxy}.

Equation \ref{eq:exoOortMass} scales from an assumed solar system Oort Cloud mass of $2\,\text{M}_{\oplus}$. Estimates of the solar system's current Oort Cloud mass are typically based on long-period comet detections and return results on the order of $M_{\text{OC}, \odot} \simeq 2\,\text{M}_{\oplus}$ \citep{Oort1950, weissman1983oort, heisler1990oortMass, boe2019PS1comets}. In the current era of survey sensitivity, this value is uncertain by a factor of up to 10 \citep{dones2004oort}. Importantly, these aforementioned values often correspond to the ``outer Oort Cloud," with semimajor axes larger than $10^{4}\,\text{au}$. \cite{hills1981cometShower} pointed out a selection effect in long-period comet counts against minor planets that reside at $10^{3}-10^{4}\,\text{au}$ since these objects would require strong perturbations to enter the inner solar system. Therefore, the inner Oort Cloud's mass could be several times larger than the outer Oort Cloud's mass with a negligible effect on long-period comet statistics \citep{dones2004oort}. The LSST will place stronger constraints on $M_{\text{OC}, \odot}$ by detecting fainter solar system long-period comets \citep{jones2009LSSTSolarSystem}. This refined knowledge of our own solar system will bolster future efforts to compare exo--Oort Clouds to our solar system's cometary inventory.

As mentioned in Section \ref{sec:intro}, theoretical models of the early solar system have proposed that an approximately $30\,M_{\oplus}$ reservoir of planetesimals from outside of Neptune's original orbit was scattered to larger semimajor axes (i.e. \citealp{Levison2008}). The mass of this initial planetesimal belt places an upper bound on the Oort Cloud, but dynamical simulations indicate that around 90\% of the original material should have been completely ejected to interstellar space; the gravitational perturbations by the giant planets are usually too powerful to lodge small bodies in the Oort Cloud region \citep{tremaine1993pulsarOortCloud, wyatt2017designOortCloud}.

\subsection{Mass Partitioning into Exo-Comet SFDs}

Given the mass budget of a star's exo--Oort ejecta in Equation \ref{eq:exoOortMass}, we next divide that reservoir into exo-comets with characteristic SFDs. We assume a characteristic single exponent power-law

\begin{equation} \label{eq:SFD}
    \frac{dN(>r)}{dr} \propto r^{-q}\,.
\end{equation}

\noindent In Equation \ref{eq:SFD}, $N(> r)$ represents the number of objects with radius greater than $r$. We denote the SFD slope with $q$. Theoretical expectations for a steady-state, collisionally-evolved population lead to $q = 3.5$, provided that material strength is constant with size \citep{dohnanyi1969cascade}. Distributions with $q > 4$ sequester most of the mass in the smallest objects, whereas SFDs with $q < 4$ exhibit the opposite behavior. For $q > 0$, the greatest number of objects are at small sizes.

We assume a universal and representative SFD for all exo--Oort Clouds regardless of the stellar mass, age, metallicity, or planetary architecture. However, the collisional physics and accretion processes that sculpt exo-comet SFDs will certainly vary from system-to-system. Piecewise broken power-law distributions with various slopes for each size regime have been historically invoked to fit solar system minor planet populations \citep{bottke2005fossilized, kenyon2008collisionalKBOs} and even hypothesized to describe the interstellar small body reservoir \citep{moro2009will, moro2018originI, moro2018originII, oumuamua2019natural}. For this study, these considerations would add unnecessarily complex parameter space to our results while diverting attention from our main objective: the possibility of LSST constraints on exo--Oort Cloud existence and structure. Notably, we consider $q$ values that are steeper than those of the solar system's minor planet populations and especially the long-period comets \citep{Bauer2017SFDcomets}.

By integrating Equation \ref{eq:SFD}, we derive the cumulative number of Jurads larger than size $r_{1}$ relative to the cumulative number of Jurads larger than size $r_{2}$ as

\begin{equation} \label{eq:SFDintegrated}
    \frac{N(>r_{1})}{N(>r_{2})} = \bigg(\frac{r_{1}}{r_{2}}\bigg)^{-q + 1}\,.
\end{equation}

\noindent With this integrated form of the SFD, we implement the following numerical procedure to estimate the population of Jurads that will be ejected by the Sun during its post--main sequence evolution.

\begin{enumerate}
    \item Choose a value of $M_{\text{J}, *}$, the total mass of Jurad ejecta, from Equation \ref{eq:exoOortMass}.
    \item Choose an assumed power-law slope $q$.
    \item Create 1000 bins for radius $r$ equidistant in log-space, ranging from chosen values of $r_{\text{min}}$ to  $r_{\text{max}}$. These minimum and maximum values are where we truncate the SFD.
    \item Compute the cumulative number $N(>r)$ of Jurads relative to $N(>r_{\text{min}})$ by repeatedly applying Equation \ref{eq:SFDintegrated}. We calculate this cumulative value for the 1001 $r$ values that correspond to the bin endpoints. For the $i^{\text{th}}$ bin, we denote these end-point radii as $r_{i,\text{min}}$ and $r_{i,\text{max}}$ for the minimum and maximum, respectively.
    \item Find the number of objects in the $i^{\text{th}}$ bin, objects with radii larger than $r_{i,\text{min}}$ and smaller than $r_{i,\text{max}}$,  by calculating
    \begin{equation} \label{eq:differentialNumber}
        N_{i}= N(>r_{i, \mathrm{min}}) - N(>r_{i, \mathrm{max}})\,,
    \end{equation}
    with the values from Step 3. After doing this subtraction for all endpoints, we have the relative number of exo-comets in each of the 1000 bins.
    \item Determine the normalization factor $f_{\text{N}}$ such that the total mass of ejected Jurads matches the original assumption for $M_{\text{J}, *}$. We compute
    \begin{equation} \label{eq:normalizationFactor}
        f_{\mathrm{N}} = M_{\mathrm{J}, *}\Bigg(\frac{4\pi}{3} \rho_{\mathrm{iso}}\sum_{i = i}^{N_{\mathrm{bins}}} r_{i}^{3}\Bigg)^{-1}\,,
    \end{equation}
    where $r_{i}$ is the radius corresponding to the mean of the endpoints from the $i^{\text{th}}$ bin and $\rho_{\text{iso}}$ is the bulk density of the small bodies. Our bin size resolution is sufficiently fine that this linear approximation is acceptable.
    \item Multiply the relative numbers of Jurads from Step 4 by the normalization factor $f_{\text{N}}$ in Equation \ref{eq:normalizationFactor} by the result from Equation \ref{eq:exoOortMass} to determine the number of Jurads ejected in each size bin.
\end{enumerate}

\begin{figure}
    \centering
    \epsscale{1.2}
    \plotone{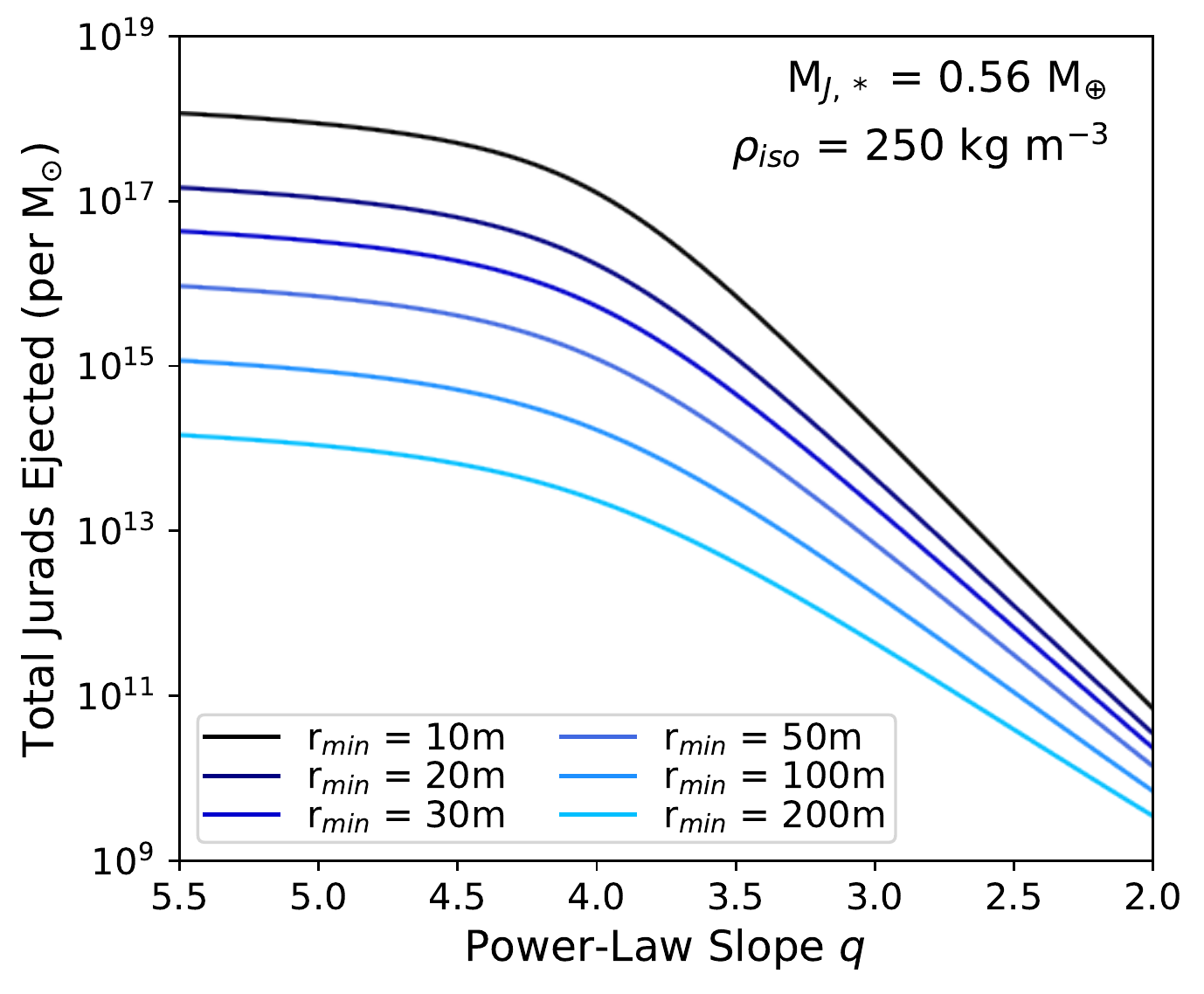}
    \caption{Total number of Jurads ejected during post--main sequence stellar evolution versus the assumed power-law slope $q$ in Equation \ref{eq:SFD}. Our results are normalized to a $1\,\msun$ star, consistent with the scaling relation in Equation \ref{eq:exoOortMass}. These solutions were computed by summing over all size bins in each SFD. We display ejection numbers for different values of $r_{\text{min}}$, the smallest size included in the population, as curves of different colors.}
    \label{fig:ejectedISOs}
\end{figure}

With this procedure, we calculated the Jurad population expected from a star with the fiducial values from Equation \ref{eq:exoOortMass} (Figure \ref{fig:ejectedISOs}). These results and can be scaled-up or scaled-down for different stellar masses, exo--Oort Cloud masses, metallicities, and ejection fractions. We evaluated a range of power-law slopes and minimum small body radii in the SFD parameterizations: $r_{\text{min}} \in \{10, 20, 30, 50, 100, 200\}\,\text{m}$ and $q \in [-5.5, 2.0]$. For each ($q$, $r_{\text{min}}$) combination, Figure \ref{fig:ejectedISOs} shows the total number of Jurads that would be generated. We verified that our final value for the total mass ejected matched our initial assumption. We have assumed that all Jurads are spherical, or equivalently, that the radius is representative of an effective value. Drawing on values for Arrokoth and small solar system long-period comets --- we refer the reader to Table S1 from \cite{keane2022arrokoth} for a collated set --- we assign bulk density $\rho_{\text{iso}} = 250\,\text{kg}\,\text{m}^{-3}$ to all Jurads regardless of physical size.

Our procedure normalizes the Jurad reservoirs to a consistent integrated mass for all SFDs, differing from \cite{moro2018originII} who anchored the distributions to $10^{12}(M_{*}/\msun)$ Jurads larger than $r = 1150\,\text{m}$. The flat slopes on our Figure \ref{fig:ejectedISOs} for $q>4$ reflects the trend for these SFDs to sequester most of the mass in objects near size $r_{\text{min}}$. At $q \simeq 5.5$, the difference in number between the SFDs with $r_{\text{min}} = 10\,\text{m}$ and $r_{\text{min}} = 100\,\text{m}$ is approximately a factor of 1000 and corresponds to the mass ratio between the smallest objects in these respective distributions. For shallow $q$, the densities begin to converge as most of the mass is sequestered in objects of $r_{\text{max}}$. The slopes steepen on Figure \ref{fig:ejectedISOs} for $q < 4$, when most of the total mass is sequestered in the largest sizes.

\subsection{Number Density in the Solar Neighborhood}

Assuming that ejection into interstellar space imparts negligible speed in addition to the stellar velocity, exo-comets in the solar neighborhood will reflect the same phase space distribution of galactic orbits as their original host stars. The collisionless Boltzmann equation should better describe the more populous interstellar small body reservoir than the stellar dynamics to which the formalism is often applied. Thus, we infer a local Jurad population from the distribution of white dwarfs.

Given the scaling of Equation \ref{eq:exoOortMass} with $M_{\text{MS}}$, the Jurad number density depends on the average main sequence mass of the host stars. To calculate this value, we first assume a power-law stellar IMF of the form

\begin{equation} \label{eq:IMF}
    \frac{dN(>M_{\text{MS}})}{dM_{\text{MS}}} \propto M_{\text{MS}}^{-\zeta}\,,
\end{equation}

\noindent where $\zeta \simeq 2.35$ \citep{salpeter1955IMF}.

We find the average $M_{\text{MS}}$ of potential Jurad progenitor stars by computing

\begin{equation} \label{eq:avgMassPMS}
    \langle M_{\text{PMS}} \rangle = \frac{\int_{1\,M_{\odot}}^{8\,M_{\odot}} M_{\mathrm{MS}} M_{\mathrm{MS}}^{-2.35} f_{\mathrm{PMS}}(M_{\mathrm{MS}})dM_{\mathrm{MS}}}{\int_{1\,M_{\odot}}^{8\,M_{\odot}}M_{\mathrm{MS}}^{-2.35} f_{\mathrm{PMS}}(M_{\mathrm{MS}})dM_{\mathrm{MS}}}\,,
\end{equation}

\noindent where $f_{\text{PMS}}(M_{\text{MS}})$ is the fraction of stars with main sequence mass $M_{\text{MS}}$ that have reached the post--main sequence on the galactic lifetime \citep{moro2018originII} and is approximated as

\begin{equation}
    f_{\mathrm{PMS}}(M_{*}) \simeq \frac{10^{10}\,\mathrm{yr} - 10^{10}\,\mathrm{yr}\,(M_{\mathrm{MS}}/\msun)^{-2.5}}{10^{10}\,\mathrm{yr}}\,.
\end{equation}

Evaluating Equation \ref{eq:avgMassPMS} gives $\langle M_{\text{PMS}} \rangle \simeq 2.6\,\msun$. Referencing the local white dwarf number density as $n_{\text{WD}} = 5.5\times10^{-3}\,\text{pc}^{-3}$ \citep{munn2017WDdensity}, it follows that the mass density of the progenitor AGB stars that we must reference is $\rho_{\text{AGB}} = 0.014\msun\,\text{pc}^{-3}$. Next, the focusing factor by which the Sun's gravitational potential enhances the number density of interstellar interlopers within the solar system is \citep{raymond2018implications}

\begin{equation} \label{eq:gravitationalFocusing}
    \xi(v_{\mathrm{iso}}) \equiv \bigg(1 + \frac{4G\mathrm{M}_{*}}{v^{2}_{\mathrm{iso}, \infty}d_{*}}\bigg)\,.
\end{equation}

\noindent In Equation \ref{eq:gravitationalFocusing}, $v_{\text{iso}, \infty}$ is the velocity of the exo-comets at infinite distance from the Sun and $d_{*}$ is the heliocentric distance at which $\xi$ is computed. We find this inbound velocity $v_{\text{iso}}, \infty$ by adding the necessary components in quadrature as

\begin{equation}\label{eq:characteristicVelocity}
    v_{\mathrm{iso}, \infty} \simeq \sqrt{\sigma_{R}^{2} + \sigma_{\phi}^{2} + \sigma_{z}^{2} + v_{\odot, \mathrm{LSR}}^{2}}\,,
\end{equation}

\noindent where the $\sigma$ values are the velocity dispersions taken from Table \ref{tab:velocityTable} and $v_{\odot, \text{LSR}} = 18\,\text{km}\,\text{s}^{-1}$ is the solar velocity versus the Local Standard of Rest \citep{schonrich2010LSR} Assuming the white dwarf kinematics from Table \ref{tab:velocityTable}, $v_{\text{iso}, \infty} \simeq 65\,\text{km}\,\text{s}$. For $d_{*} = 1\,\text{au}$, we find a modest density enhancement of $\xi(65\,\text{km}\,\text{s}^{-1}) \simeq 1.8$. In comparison, $\xi(27\,\text{km}\,\text{s}^{-1}) \simeq 6$ for \om-like velocities.

\begin{table}[]
\centering
\caption{Representative parameters for the velocity distributions of planetary nebulae and white dwarfs in the Milky Way from \cite{delhaye1965solarMotion}: the asymmetric drift velocity $v_{\text{A}}$, and the radial, azimuthal, and vertical velocity dispersions $\sigma_{R}$, $\sigma_{\phi}$, and $\sigma_{z}$, respectively.}
\begin{tabular}{|c|c|c|}
\hline
\textbf{} & \textbf{Planetary Nebulae} & \textbf{White Dwarfs} \\ \hline $v_{\text{A}}$ & $24\,\text{km}\,\text{s}^{-1}$ & $32\,\text{km}\,\text{s}^{-1}$ \\ \hline
$\sigma_{R}$ & $45\,\text{km}\,\text{s}^{-1}$ & $50\,\text{km}\,\text{s}^{-1}$ \\ \hline
$\sigma_{\phi}$ & $35\,\text{km}\,\text{s}^{-1}$ & $30\,\text{km}\,\text{s}^{-1}$ \\ \hline
$\sigma_{z}$ & $20\,\text{km}\,\text{s}^{-1}$ & $25\,\text{km}\,\text{s}^{-1}$ \\ \hline
\end{tabular}
\label{tab:velocityTable}
\end{table}

Therefore, we can estimate the number density of Jurads in the solar system as

\begin{equation} \label{eq:localNumberDensity}
    \begin{split}
    n_{\mathrm{J}} = 1\times10^{-2}\,\mathrm{au}^{-3} \bigg(\frac{n_{\mathrm{WD}}}{5.5\times10^{-3}\,\mathrm{pc}^{-3}}\bigg)\\\,\bigg(\frac{\xi(v_{\mathrm{iso}})}{1.8}\bigg)\,\bigg(\frac{N_{\mathrm{ej}}}{10^{16}}\bigg)\,,
    \end{split}
\end{equation}

\noindent where $N_{\text{ej}}$ is the number of ejected Jurads ejected from the typical white dwarf.

This fiducial $n_{\text{J}}$ has been estimated from scaling the results on Figure \ref{fig:ejectedISOs}. While this value of order $10^{-2}\,\text{au}^{-3}$ for $r \gtrsim 30\,\text{m}$ objects and steep ($q \gtrsim 4.5$) power-law slopes is an order-of-magnitude smaller than the estimated number density for \om-like interlopers, the discovery of Jurads still could be possible for a few reasons:

\begin{enumerate}
    \item The LSST's limiting brightness for discovering NEOs will be 2-3 magnitudes fainter than previous wide-field campaigns \citep{ivezic2019lsst}, expanding the search volume for Jurads.
    \item Faster inbound velocities for Jurads increase the collision rate of this population with the inner solar system versus what would be realized for \om-like kinematics.
\end{enumerate}

For these reasons, we conduct a more detailed assessment of the LSST's ability to discover Jurad interlopers.

\section{LSST Sensitivity to Jurads} \label{sec:LSST}

With the size-dependent number densities of Jurads in nearby interstellar space, our next aim is to estimate the total search volume for which the Survey is sensitive to these interlopers. We implement and compare three models of the LSST's search volume to assess the prospects for detecting Jurads. Despite the varying complexity of the models that we will use -- the procedures and their explanations are ordered by increasing computational runtime -- the results agree at the order-of-magnitude level for $r \gtrsim 70\,\text{m}$ interlopers. Given that the total masses and SFDs of exo--Oort Clouds are altogether unconstrained, our model of the LSST does not dominate the uncertainty in Jurad detection rates.

\subsection{Model \#1: Representative Quarter-Sphere Volume}

Our goal is to calculate the LSST's total search volume $V_{\text{tot}}(H)$ over the nominal ten-year survey for interlopers with absolute magnitude $H$. Converting $H$ to an interloper radius will give the necessary Jurad number density $n_{\text{J}}(r)$ such that the LSST would be expected to find one interloper satisfying this $H$ criterion.

Inactive objects will have absolute magnitudes that are well-approximated by the following formula given by \citep{russell1916albedo}:

\begin{equation} \label{eq:absoluteMagnitude}
    H(r_{\mathrm{km}}, p) = \frac{\log_{10}(2r_{\mathrm{km}}) + 0.5\log_{10}(p) - 3.1236}{-0.2}\,.
\end{equation}

\noindent In Equation \ref{eq:absoluteMagnitude}, $r_{\text{km}} = r(1\,\text{km})^{-1}$ and $p$ is the geometric albedo. The absolute magnitude $H$ is the brightness of a minor planet viewed face-on at a heliocentric distance of $1\,\text{au}$. This construct corresponds to the (impossible) viewing position of the Sun's center but provides a useful value that is intrinsic to a given small body.

We assume that Jurads are inactive but note that this ansatz will underappreciate the LSST's detection statistics if these processed exo-comets do have bright coma. We assume constant $p = 0.06$, corresponding to asteroid-like values. At a given point in the solar system, we can translate an interloper's absolute magnitude into an apparent visual magnitude $m_{\text{V}}$ as \citep{bowell1989asteroidsII}

\begin{equation} \label{eq:apparentMagnitude}
    m_{\mathrm{V}} = H + 2.5\log_{10}\Bigg(\frac{\Delta_{\odot}^{2}\Delta_{\oplus}^{2}}{\gamma(\alpha)}\Bigg)\,,
\end{equation}

\noindent where $\Delta_{\odot}$ and $\Delta_{\oplus}$ are the heliocentric and geocentric distances in au, respectively. The photometric phase correction $\gamma(\alpha)$ accounts for the viewing geometry of small bodies and is approximated via

\begin{equation} \label{eq:gammaPhotometric}
    \gamma(\alpha) = (1-G)\Phi_{1}(\alpha) + G\Phi_{2}(\alpha)\,.
\end{equation}

\noindent In Equation \ref{eq:gammaPhotometric}, the basis functions for $\gamma(\alpha)$ are

\begin{equation} \label{eq:phaseBasis}
\begin{cases}
    \Phi_{1}(\alpha) \simeq \exp\big[-3.33\,\tan(0.5\alpha)^{0.63}\big]\\
    \Phi_{2}(\alpha) \simeq \exp\big[-1.87\,\tan(0.5\alpha)^{1.22}\big]\,.
    \end{cases}
\end{equation}

\noindent In Equation \ref{eq:phaseBasis}, the phase angle $\alpha$ is the Sun-interloper-Earth angle during the epoch of observation. The coefficients have been optimized over aggregate asteroid photometry by \cite{muinonen2010phaseFunction}.

For this first LSST model, we set $\alpha = 50^{\circ}$ since this value corresponds to the average phase angle of objects detected by ATLAS with heliocentric distances less than $1.3\,\text{au}$ \citep{Tonry2018-ATLAS}. A better estimate for Jurad phase angles would require injection-recovery simulations with the LSST discovery pipeline and is beyond the scope of this study. ATLAS covers the nighttime sky every few days, providing a reasonable empirical proxy for the phase angle distribution for the LSST's detections of NEOs. With this assumption, we will calculate the distance from Earth at which an inactive small body could be discovered at this phase angle by the LSST. We then use this value to calculate a representative search volume in the quarter-sphere probed by the Survey. The heliocentric distance is given by

\begin{equation} \label{eq:heliocentricDistance}
    \Delta_{\odot} = \Delta_{\oplus}\cos(\alpha) + \sqrt{(1\,\mathrm{au})^{2} - \Delta_{\oplus}^2 + \Delta_{\oplus}^{2} \cos(\alpha)}\,.
\end{equation}

\noindent We can substitute $\Delta_{\odot}$ from Equation \ref{eq:heliocentricDistance} into Equation \ref{eq:apparentMagnitude} to determine the $\Delta_{\oplus}$ at which objects of size $r$ will appear at the LSST V-band magnitude limit of $m_{\text{V}} \simeq 24\,\text{mag}$. We use the \texttt{fsolve} functionality in \texttt{scipy} \citep{virtanen2020scipy} and validate our results by numerically solving in Mathematica.

Next, we must quantify limitations on the LSST's ability to detect fast-moving interstellar interlopers. First, objects with $m_{\text{V}} < 16\,\text{mag}$ will saturate the detector and will likely not be detected \citep{ivezic2019lsst}. Second, objects moving with angular velocities larger than $\omega_{\text{lim}} \gtrsim 10^{\circ}\,\text{d}^{-1}$ will either spread their photons across too many pixels to be identified or move too far between frames for orbits to be properly recovered by tracklet-building algorithms. \cite{veresChesley2017LSST} estimated that these ``trailing losses" for minor planets moving at this angular velocity are equivalent to decreasing the brightness by two apparent magnitudes. Third, the LSST will require that objects be detected on three separate nights to be discovered. With a typical revisit time of $3\,\text{days}$ for a given field \citep{ivezic2019lsst}, objects must remain in the search volume for more than $1\,\text{wk}$ to be identified. Conservatively, we take $10\,\text{d}$ to be the minimum residence time in the search volume for discoverable interlopers.

Each of these considerations causes an effective reduction of the instantaneous search volume $V_{\text{inst}}(H)$ for objects of a given size. In the context of our model, these detection restrictions set a minimum $\Delta_{\oplus}$ for which a Jurad could be observed. Thus, the instantaneous search volume is

\begin{equation} \label{eq:instantaneousVolume}
    V_{\mathrm{inst},1}(H) = \Big[\frac{\pi}{3}\Big]\big[R_{\mathrm{eff},1}(H)^{3} - R_{\mathrm{lim}}(H)^{3}\big]\,,
\end{equation}

\noindent where the ``effective search radius" $R_{\text{eff},1}(H)$ is the geocentric distance at which an object of absolute magnitude $H$ appears at $m_{\text{V}} = 24\,\text{mag}$ at phase angle $\alpha = 50^{\circ}$ and $R_{\text{lim}}$ is an estimated minimum geocentric distance for the interloper to be detectable. For this model of the LSST's capabilities, we conservatively assume this limiting geocentric distance to be

\begin{equation} \label{eq:limitingRadius}
    R_{\mathrm{lim}}(H) = \max\Big\{R_{\mathrm{sat}}(H), R_{\mathrm{tr}}, R_{10\,\mathrm{d}}\Big\}\,,
\end{equation}

\noindent where $R_{\text{sat}}$, $R_{\text{tr}}$, and $R_{10\,\text{d}}$ are the limiting geocentric detection distances due to saturation, trailing, and search strategy effects, respectively. 

We calculate $R_{\text{sat}}(H)$ in an analogous manner as we did to find $R_{\text{eff}}$ but changed the small body limiting brightness to $m_{\text{V}} = 16\,\text{mag}$. To find $R_{\text{tr}}$, we use the small angle approximation to obtain the minimum $\Delta_{\oplus}$ to guarantee that an interloper will not exceed the LSST's angular velocity limit as

\begin{equation} \label{eq:streakingRadius}
    R_{\mathrm{tr}}\simeq 0.27\,\text{au} \bigg(\frac{v_{\text{iso},\oplus}}{82\,\text{km}\,\text{s}^{-1}}\bigg)\bigg(\frac{\omega_{\text{lim}}}{10^{\circ}\,\mathrm{d}^{-1}}\bigg)^{-1}\,,
\end{equation}

\noindent where $v_{\text{iso},\oplus}$ is the velocity of the interstellar object with respect to the Earth at $\Delta_{\odot} = 1\,\text{au}$. To calculate this relative velocity, we first calculate the heliocentric velocity of the interloper from energy conservation as

\begin{equation} \label{eq:velocityInterlopers}
    \frac{-GM_{\odot}}{R} + \frac{v_{\mathrm{iso},\odot}(R)^{2}}{2} = \frac{v_{\mathrm{iso}, \infty}^{2}}{2}\,,
\end{equation}

\noindent where $R$ is the instantaneous heliocentric distance. Substituting $R = 1\,\text{au}$, we find $v_{\text{iso},\odot}(1\,\text{au}) = 77\,\text{km}\,\text{s}^{-1}$. Then, we find a representative geocentric velocity as

\begin{equation} \label{eq:geocentricVelocity}
    v_{\text{iso},\oplus} = \sqrt{v_{\mathrm{iso},\odot}(1\,\mathrm{au})^{2} + v_{\mathrm{orb},\oplus}^{2}} \simeq 82\,\mathrm{km}\,\mathrm{s}^{-1}\,,
\end{equation}

\noindent where $v_{\text{orb},\oplus} \simeq 30\,\text{km}\,\text{s}^{-1}$ is the orbital velocity of the Earth around the Sun. For reference, $R_{\text{tr}} \simeq 0.16\,\text{au}$ for \om-like trajectories.

Here, we have considered interlopers that pass by Earth tangent to the sky-plane. Since exo-comets with steeper approach angles will have lower angular velocities, our assumption will yield a conservative estimate for the LSST search volume.  Properly accounting for the detectability of small bodies moving with high angular velocities would require detailed injection-recovery simulations using a framework like the Moving Object Processing System (MOPS) discovery pipeline \citep{denneau2013MOPS} and synthetic LSST frames \citep{veresChesley2017LSST}. Interestingly, Equation \ref{eq:streakingRadius} indicates that \om{} may have gone undetected by Pan-STARRS1 during its closest geocentric approach if this interloper had exhibited Jurad-like kinematics. 

Finally, we compute $R_{\text{10}\,\text{d}}$ as the distance traversed over ten days by an interloper moving with velocity $v_{\text{iso},\oplus} \simeq 82\,\text{km}\,\text{s}^{-1}$ and find $R_{10\,\text{d}} \simeq 0.47\,\text{au}$. With these values as inputs to Equation \ref{eq:limitingRadius}, we compute $V_{\text{inst}}(H)$ for objects of arbitrary $H$ in Equation \ref{eq:instantaneousVolume}. To compute the total survey volume for Model \#1, $V_{\text{LSST},1}(H)$, we estimate the number of times that a new set of interlopers occupies the LSST's search volume during the ten-year campaign \citep{moro2018originII}. At the order-of-magnitude level, we can define a ``refresh time" $t_{\text{ref}}$ as the time required for interlopers to pass through the search volume. Because the LSST will revisit fields approximately every 3 nights \citep{ivezic2019lsst} --- with actual values dependent on a finalized cadence and weather --- we define this characteristic timescale as

\begin{equation} \label{eq:refreshTime1}
    t_{\text{ref},1}(H) \approx \max\bigg[10\,\mathrm{d}, \frac{R_{\mathrm{eff},1}(H)}{v_{\mathrm{iso},\oplus}}\bigg]\,.
\end{equation}

\noindent Note that our aforementioned $R_{10\,\text{d}}$ criterion implies that $R_{\text{eff}}/v_{\text{iso}, \oplus} > 10\,\text{d}$ in all relevant cases.

\cite{hoover2022population} accounted for the typical residence time of interlopers in the LSST's search volume by multiplying the analog of their $R_{\text{eff},1}(H)/v_{\text{iso}, \oplus}$ term by 4/3. For this model, we do not incorporate a similar factor because of the asphericity of the LSST's search volume. Nonetheless, the analytic work from \cite{hoover2022population} indicates that any multiplicative factor is likely close to unity and unlikely to affect our results at order-of-magnitude precision. 

From Equations \ref{eq:instantaneousVolume} \& \ref{eq:refreshTime1}, we determine the total search volume for objects of a given $H$ as

\begin{equation} \label{eq:model1Result}
    V_{\mathrm{tot},1}(H) = \bigg(\frac{10\,\mathrm{yr}}{t_{\mathrm{ref},1}(H)}\bigg) \bigg(V_{\mathrm{inst},1}(H)\bigg)\,.
\end{equation}

For Model \#1, we have ignored considerations of weather \citep{ivezic2019lsst}, interloper shape \citep{levine2023shape}, and seasonality \citep{Seligman2018}, among other minor effects.

\subsection{Model \#2: Numerically-Integrated Search Volume}

To validate Model \#1's approximation of a quarter-sphere search volume, we next numerically determined the instantaneous LSST search volume for objects with absolute magnitude $H$ without the assumption of a representative phase angle. The true shape of the LSST's search volume is cone-like, since minor planets are brightest at opposition. We find the maximum geocentric distance $R_{\text{max}}(H)$ for which the interloper could be found at $m_{\text{V}} = 24\,\text{mag}$ by setting $\alpha = 0$ and solving for $\Delta_{\oplus}$ in Equation \ref{eq:apparentMagnitude}. Next, we divide the volume encompassed by this geocentric distance into spherical coordinate grid cells as $R \in \{0, R_{\text{max}}(H)\}$, $\phi \in \{\pi/2, \pi\}$, $\theta \in \{-\pi/2, \pi/2\}$. Then, we find the instantaneous search volume by evaluating

\begin{equation} \label{eq:integrateVolume}
    V_{\mathrm{inst},2}(H) = \int_{R_{1}}^{R_{2}}\int_{\phi_{1}}^{\phi_{2}}\int_{\theta_{1}}^{\theta_{2}}\mathcal{B}(R, \theta, \phi) R^{2} \sin(\phi) d\theta\,d\phi\,dR\,,
\end{equation}

\noindent via the midpoint method, where $\mathcal{B}(R, \theta, \phi)$ is a Boolean that is \texttt{True} if the interloper is detectable in the volume element and \texttt{False} is the object is invisible to the LSST in the grid cell. The bounds on this triple integral are given by the aforementioned limits. This truth value encompasses the three considerations from Model \#1 encapsulated by $R_{\text{sat}}(H)$, $R_{\text{tr}}$, and $R_{10\,\text{d}}$.

From the instantaneous search volume given by Equation \ref{eq:integrateVolume}, we can calculate the refresh time as

\begin{equation}\label{eq:refreshTime2}
    t_{\mathrm{ref},2}(H) = \max\bigg[10\,\mathrm{d},\;\frac{(3V_{\mathrm{inst},2}(H))^{1/3}}{\pi v_{\mathrm{iso},\oplus}}\bigg]\,,
\end{equation}

\noindent and apply an analogous version of Equation \ref{eq:model1Result} for $t_{\text{ref},2}$ and $V_{\text{inst},2}(H)$ to determine $V_{\text{tot},2}(H)$ for Model \#2. 

\begin{figure}
    \centering
    \epsscale{1.2}
    \plotone{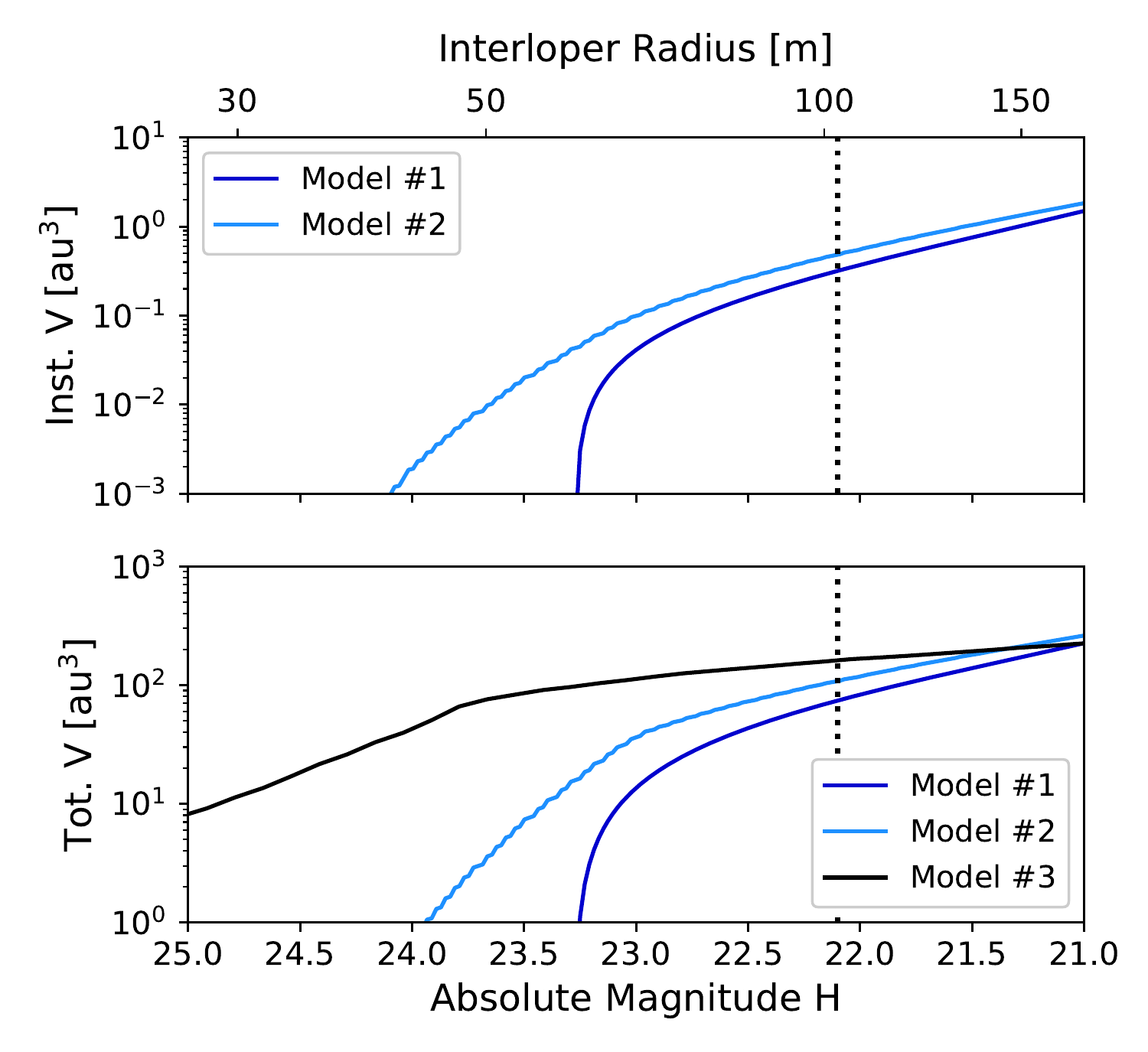}
    \caption{Comparison of our three methods of calculating the LSST's search volume in which the Survey could detect Jurads. The top panels shows values of instantaneous search volume $V_{\text{inst}}(H)$, while the lower panel shows the total volume $V_{\text{LSST}}$ over the ten-year survey. ``Model \#1" corresponds to the representative quarter-sphere formulation, ``Model \#2" shows the result of numerically integrating the LSST search volume, and ``Model \#3" shows the application of simulations based on those by \citet{hoover2022population} to this situation. Only Model \#1 and Model \#2 return values for instantaneous volume. For reference, the size of \om{} is marked by the vertical dotted line.}
    \label{fig:compareLSSTmodels}
\end{figure}

We compare the results of the instantaneous volume estimates on the top panel of Figure \ref{fig:compareLSSTmodels}, finding order-of-magnitude agreement for Jurads with sizes $r \gtrsim 70\,\text{m}$. Recently, \cite{ezell2023detectionLSST} calculated the detection rate of $r < 50\,\text{m}$ interstellar objects. Although trailing loss was acknowledged, this effect was not incorporated into their analysis. Our results show that trailing loss plays a significant role in detecting small interlopers. The increasing asphericity of the LSST's search volume at high $H$ likely accounts for the divergence of our Model \#1 and Model \#2 for small interlopers.

While the Rubin Observatory will spend 90\% of the night executing the LSST, the remaining 10\% will be devoted to other science programs \citep{schwamb2023LSSTcadence}. These initiatives may include deep-drilling fields and/or a near-Sun survey. We have not considered the prospects for detecting exo-comets in these additional images, and we defer an assessment of alternate search strategies for interstellar small body discoveries to future work.

\subsection{Model \#3: \textit{N}-Body Dynamical Simulations}

\cite{hoover2022population} modeled the number of interlopers within the LSST's search volume by numerically resolving a set of interloper trajectories with inbound kinematics that were representative of stars in the solar neighborhood. Here, we assess the LSST's ability to detect Jurads by applying the same numerical scheme while updating the inbound velocity distributions to the values from Table \ref{tab:velocityTable}. Notably, this framework returns the number of ``detectable" objects and not the actual number of interlopers that would be discovered by the LSST. Trailing losses, the survey cadence, and the requirement that an object be visible over three different telescope visit nights for linking will each reduce the number of ``detected" objects from the number of ``detectable" objects reported by \cite{hoover2022population}. Nonetheless, assessing the detectability of characteristic interloper trajectories through the solar system provides an important benchmark for our other two models.

To be counted as detectable, \citet{hoover2022population} required that an interloper to meet the detection criteria for a single integration timestep of approximately $1\,\text{d}$. For the simulations in our Model \#3, interlopers must satisfy the LSST's brightness requirements and remain in the field-of-view for at least \textit{three} integration timesteps. This modification results in lower detection rates than were found by \citet{hoover2022population}, but is still less stringent than the 10d requirement that we used for Model \#1 and \#2. In addition, we consider albedo $p = 0.06$ small bodies to be consistent with the previous models.

The output of Model \#3 gives the fraction $f_{3}(r)$ of objects with size $r$ that are detectable by the LSST in a $5\,\text{au}$ sphere around the Sun. Since we find the detectability of $10^{5}$ objects that pass through this region, we can calculate the requisite interstellar number density to detect one interloper over the ten-year LSST as 

\begin{equation} \label{eq:numberDensity3}
    n_{3}(r) = \frac{9.2\times 10^{-5}}{f_{3}(r)}\,\mathrm{au}^{-3}\,.
\end{equation}

\noindent In Equation \ref{eq:numberDensity3}, the coefficient is from adapting Equation 13 from \cite{hoover2022population} to $v_{\infty} = 65\,\text{km}\,\text{s}^{-1}$.

We invert Equation \ref{eq:numberDensity3} to get $V_{\text{tot}, 3}(H) = n_{3}^{-1}(r)$, a total search volume for the LSST as calculated by this model (where we have used $r$ and $H$ interchangeably). We plot our results on the bottom panel of Figure \ref{fig:compareLSSTmodels}, for comparison with Models \#1 \& \#2. Model \#3 for the ``detectable" objects diverges from Models \#1 \& \#2 for the smallest objects that we have considered. Any $r\sim30\,\text{m}$ interlopers must be close to the Earth to be seen, which necessarily results in high angular velocities and a propensity for trailing losses to occur. Knowing that Model \#3 should provide an upper bound, we will adopt results for the total LSST volume from Model \#2 ($V_{\text{tot}, 2}(H)$) for the remainder of this study.

\section{LSST Constraints on Exo--Oort Clouds} \label{sec:constraints}

Given the previous calculations on the LSST's sensitivity towards Jurads, we examine the constraints on exo--Oort Cloud occurrence and structure that could be ascertained by the Survey.

\subsection{Detecting Jurads from Model SFDs}

From Section \ref{sec:occurrence}, we have a table of $n_{\text{iso}}(r)$ in bins with endpoints $r_{i, \text{min}}$ and $r_{i, \text{max}}$. From Section \ref{sec:LSST}, we have a table of $V_{\text{tot},2}(H)$ which can be converted into a function of $r$ as $V_{\text{tot},2}(r)$ through Equation \ref{eq:absoluteMagnitude}. In order to compute population-integrated discovery expectations for the LSST, we must map the values of $V_{\text{tot},2}(r)$ to the $n_{\text{iso}}(r)$ bins. Since each of our bins spans a small size range, we linearly interpolate as 

\begin{equation} \label{eq:interpolation}
    V(r_{i, \text{min}}) = \frac{(r_{i}-r_{0})(V(r_{1})-V(r_{0}))}{r_{1}-r_{0}} + V(r_{0})\,,
\end{equation}

\noindent where $r_{0}$ and $r_{1}$ are the $r$ values in the $V_{\text{tot},2}(r)$ table that are closest to the bin endpoint $r_{i,\text{min}}$ in the $n_{\text{iso}}(r)$ table on the smaller and larger side, respectively. 

We performed a similar calculation to find $V(r_{i,\text{max}})$, then take the average of the two search volumes as the total LSST search volume for that SFD bin $V(r_{i})$. With this standardized table, we computed the expected number of discoveries within a given size bin as

\begin{equation}\label{eq:expectedLSSTbyBin}
    N_{\mathrm{LSST}}(r_{i}) = V(r_{i}) n_{\text{iso}}(r_{i})\,,
\end{equation}

\noindent where $n_{\text{iso}}(r_{i})$ is number density of Jurads in the solar neighborhood in the $i^{\text{th}}$ size bin.

By summing the results from each bin, we derived cumulative values for LSST discoveries of Jurads larger than radius $r$. Doing this summation for all bins, we find the expected total number of Jurad discoveries by the Survey as a function of the SFD inputs as

\begin{equation} \label{eq:expectedLSSTdiscoveries}
    N_{\mathrm{LSST}}(q, r_{\mathrm{min}}) = \sum_{r= r_{\mathrm{min}}}^{r_{\mathrm{max}}} n_{\mathrm{iso}}(q, r)V_{\mathrm{LSST}}(r)\,,
\end{equation}

\noindent where $r_{\text{min}}$ is the minimum radius of objects in the SFD, $r_{\text{max}}$ is the maximum radius of objects under consideration, $n_{\text{iso}}(q, r)$ is the number density of objects with radius $r$ in the solar neighborhood for the given SFD parameterization, and $V_{\text{LSST}}(r)$ is the total LSST search volume for the interlopers in a bin with mean radius $r$. In Equation \ref{eq:expectedLSSTdiscoveries}, the summation is over all bins encompassing the given size range.

\begin{figure}
    \centering
    \epsscale{1.2}
    \plotone{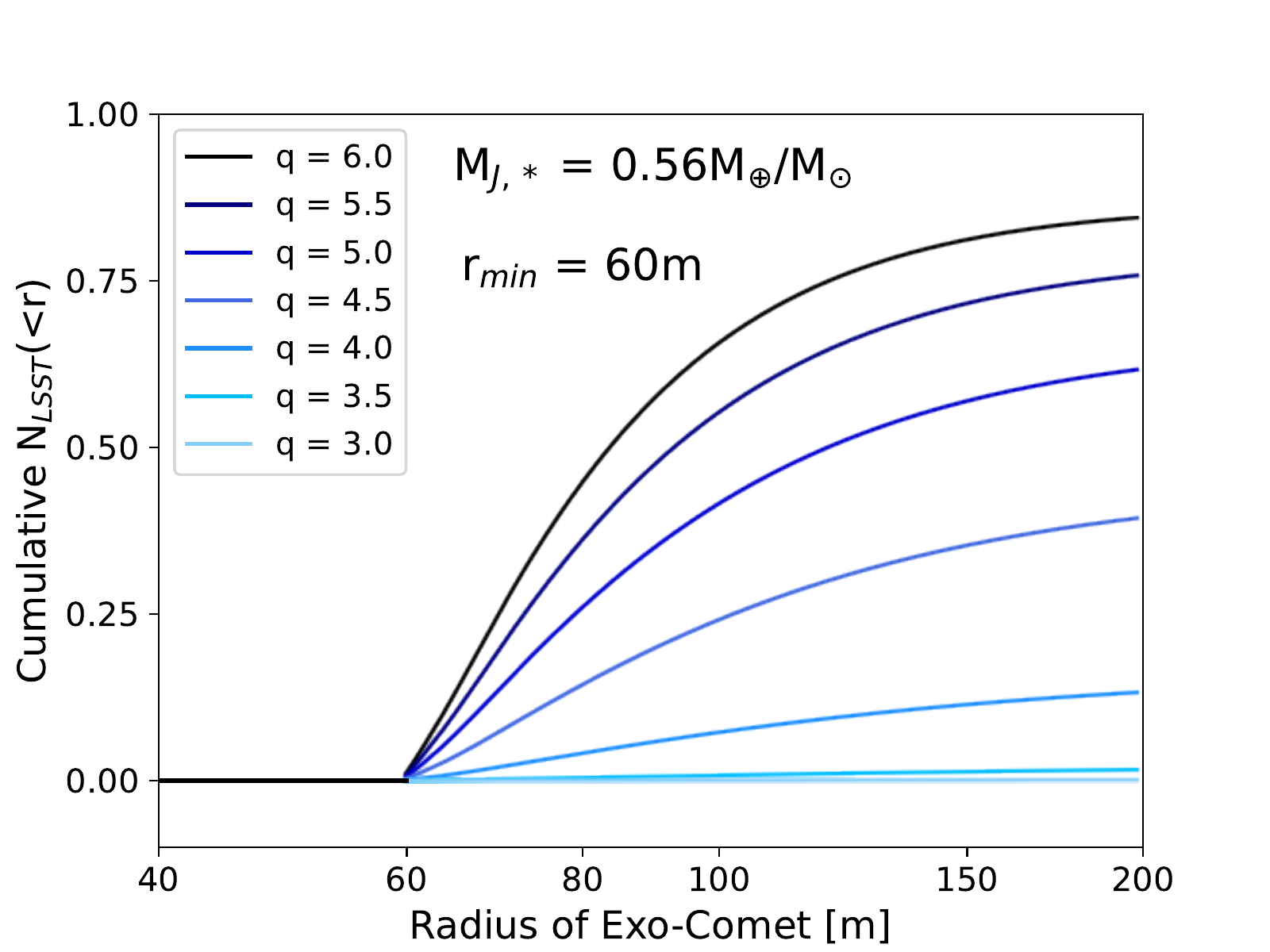}
    \caption{Expected number of LSST detections for Jurads smaller than radius $r$ versus exo-comet physical size, expressed as a cumulative value $N_{\text{LSST}}(<r)$. For this plot, we choose $r_{\text{min}} = 60\,\text{m}$ in the SFD and display the results for various power-law slopes $q$.}
    \label{fig:lsstDetections}
\end{figure}

Figure \ref{fig:lsstDetections} shows the cumulative number of expected LSST discoveries for different $q$ values with the example $r_{\text{min}} = 60\,\text{m}$. We have assumed a fiducial Jurad generation of $M_{\text{J},*} = 0.56\,\text{M}_{\oplus}/\msun$ from Equation \ref{eq:exoOortMass}, and we plot the results as the number of detections for objects larger than a given radius. Objects close to the minimum size in the SFD are the most likely to be detected, and few discoveries are expected of Jurads larger than $r\sim100\,\text{m}$ for any $q$. Steep power-laws that generate the most small objects lead to more favorable prospects for the LSST to detect a Jurad, so long as those objects are not too small for trailing losses to dominate.

\subsection{Exo--Oort Cloud Parameter Space}

\begin{figure}
    \centering
    \epsscale{1.2}
    \plotone{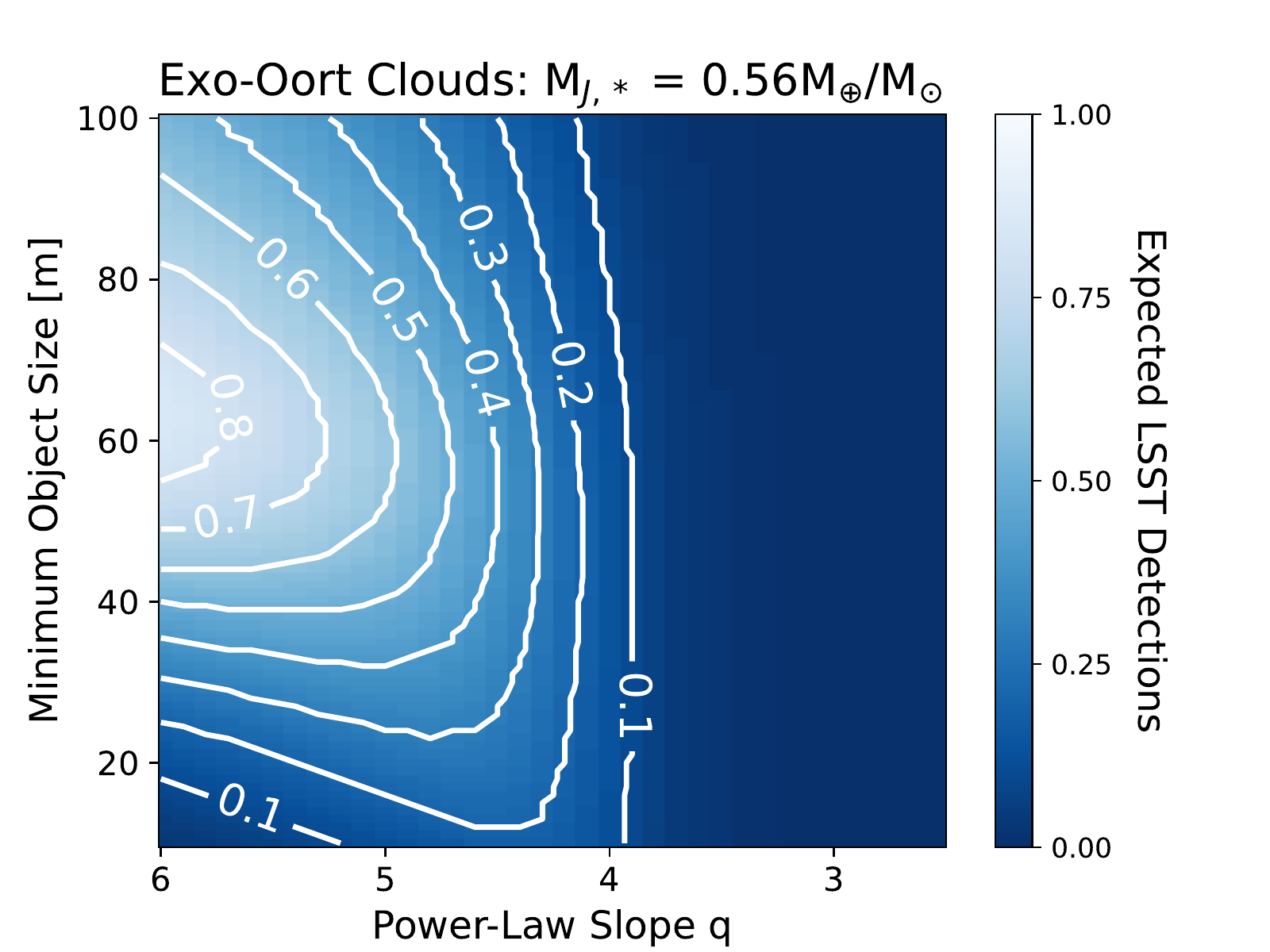}
    \caption{Expected total number of LSST detections for Jurads of all sizes, given the $r_{\text{min}}$ and $q$ of the SFD. In these calculations, we have assumed the fiducial values for exo--Oort Cloud populations from Section \ref{sec:occurrence} and the second LSST detection model from Section \ref{sec:LSST}.}
    \label{fig:lsstParameterSpace}
\end{figure}

By applying the method used to generate Figure \ref{fig:lsstDetections} to other $r_{\text{min}}$, we can interrogate the exo--Oort Cloud parameters that the LSST will constrain. Specifically, we continue to assume $M_{\text{J}, *} = 0.56\,\text{M}_{\oplus}/\msun$ and initialize a grid of SFDs in ($q$, $r_{\text{min}}$) space. We then calculate the total number of expected Jurad discoveries by the LSST for radii up to $500\,\text{m}$ via the procedure leading to Equation \ref{eq:expectedLSSTdiscoveries}. Our results are plotted on Figure \ref{fig:lsstParameterSpace}, which conveniently provides a number of insights into the LSST's ability to probe exo--Oort Cloud populations.

The number of Jurad detections by the LSST is degenerate with $r_{\text{min}}$ and $q$ in the SFDs for a given $M_{\text{J}, *}$. Nonetheless, the identification of one interloper by the LSST that is subsequently determined to be a Jurad (per the forthcoming discussion in Section 7) would lead to these broad insights on exo--Oort Clouds:

\begin{itemize}
    \item The smallest radius $r_{\text{min}}$ of Jurads is probably close to the Jurad's size itself. This conclusion comes from the results of Figure \ref{fig:lsstDetections}.
    \item Typical power-law slopes are likely to be $q > 4$. Jurads are unlikely to be discovered if $q < 4$, even in the case where $M_{\text{J}, *}$ is an order-of-magnitude larger than our fiducial assumption. Therefore, the detection of a single object of this class would immediately indicate that most of the mass in exo--Oort Clouds resides in decameter-sized objects. The SFD slopes and characteristic sizes that would be required for the LSST to detect a Jurad are strikingly different than the parameters ($q \simeq 1$, $r \simeq 1\,\text{km}$) that have been derived by missions such as \textit{NEOWISE} for solar system long-period comets \citep{Bauer2017SFDcomets}.
    \item Solid masses in the exo-Oort regions surrounding main sequence stars with A, F, and G main-sequence spectral types must be comparable to that of our outer solar system. Otherwise, the discovery of Jurads would be improbable for any combination of ($q$, $r_{\text{min}}$). 
\end{itemize}

The number densities of Jurads in interstellar space depends linearly on $M_{\text{J},*}$, so the LSST expectations in Figure \ref{fig:lsstParameterSpace} can be scaled accordingly with this assumption. Although SFDs with $r_{\text{min}} \lesssim 40\text{m}$ will generate a larger total number of objects than distributions with larger $r_{\text{min}}$, the LSST's ability to discover such populations is hampered by trailing losses and the $3\,\text{d}$ revisit time. Small Jurads require close geocentric distances to be discovered but would be fast-moving, the combination of which leads to disfavorable detection statistics.

In the event of an LSST nondetection, any of these conclusions about exo--Oort Clouds could be the cause:

\begin{itemize}
    \item The total mass sequestered in exo--Oort comets is smaller than the inferred reservoir in our solar system, regardless of the characteristic SFDs.
    \item Most of the mass in exo--Oort Clouds is in objects either larger or smaller than the decameter scale, either due to shallow power-law slopes or from the typical value of $r_{\text{min}}$.
    \item Exo-comets are (i) somehow destroyed in post--main sequence systems, (ii) not released during the AGB thermal pulses, or (iii) ejected by Galactic tides and stellar encounters at earlier times.
\end{itemize}

In total, the LSST's unprecedented search volume could yield the first direct detection of a \textit{bona fide} post--main sequence exo-comet and complement ongoing observational research on white dwarf systems. Future studies in the LSST era could benefit from a Bayesian analysis that updates priors from planet formation theory based on a Jurad detection or non-detection. Pairing such a model with injection/recovery simulations on LSST images would yield better estimates of the search volume, leading to tighter constraints on the interstellar Jurad reservoir. A Bayesian approach would offer a more nuanced understanding of how a decade of Rubin/LSST observations influences prevailing hypotheses but is beyond the aim of the current study: determining the LSST's ability to identify a Jurad. One recent example of a Bayesian method that would be useful in the context of a future Jurad discovery is the publication by \citep{flekkoy2023statisticsOumuamua}, which examined the most likely detection rate of \om-like objects.

\subsection{Relative Counts of Jurads to Other Interlopers}

Thus far, we have focused our work on the absolute counts of Jurads in the LSST since we are interested in the detectability of this population. Provided that the physical and chemical processing described in Section \ref{sec:processing} impart signatures onto Jurads that distinguish these objects from their counterparts ejected on the (pre--)main sequence, the relative count of Jurad-like to non-Jurad interlopers could be an alternative diagnostic on the occurrence of exo-Oort Clouds. Considerations of survey strategy and limiting magnitude may be less consequential for relative counts than for absolute counts, making the former measure a more robust metric.

If models of our solar system's dynamical evolution (i.e. \citealp{Levison2008}) apply more broadly to extrasolar systems, then about 10\% of the galaxy's interstellar small bodies should be Jurads. We began with such an ansatz in Equation \ref{eq:exoOortMass} for Jurad ejecta mass and the ensuing calculations. Nonetheless, we caution that a number of factors affecting Jurad counts do not affect searches for interlopers originating from main sequence stars to the same degree. Faster expected inbound velocities of Jurads increase trailing losses while dampening the concentration effect of gravitational focusing. Provided that Jurads do not have brightness-boosting tails, then these processed objects would be dimmer than Borisov-like interlopers.

Nonetheless, a na\"ive extrapolation from the Nice Model agrees at the order-of-magnitude level with our predictions in Figures \ref{fig:lsstDetections} \& \ref{fig:lsstParameterSpace}. This result indicates that selection effects against Jurads do not substantially inhibit LSST's detection capabilities versus slower-moving populations of exo-comets. We find that the LSST could see $\mathcal{O}(0.1-1)$ Jurads provided a favorable distribution of mass into decameter-scale objects. Previous work has suggested that the LSST could see $\mathcal{O}(10)$ \om-like objects \citep{cook2016realistic, hoover2022population}. Therefore, the relative number of Jurads to main sequence interlopers discovered by the LSST could illuminate the ubiquity of dynamical histories like that of our solar system.

\section{Discussion} \label{sec:discussion}

The LSST is the first observational campaign that will place meaningful constraints on exo--Oort Clouds because of its larger search volume versus those of Pan-STARRS1, the Catalina Sky Survey, and other previous efforts to discover minor planets at-scale. The (non)detection of Jurads will eliminate some astrophysically feasible scenarios of exo--Oort Cloud occurrence and lead to corresponding constraints on the formation and evolution of extrasolar systems. If the LSST makes a serendipitous Jurad discovery, then orthogonal and complementary insight will be gained towards ongoing efforts to characterize exoplanets.

\subsection{Jurads in the Context of Planet Formation}

The solar system's Oort Cloud was likely populated via perturbations from giant planets. The protosolar nebula did not have sufficient gas densities to form planetesimals at Oort Cloud distances. Therefore, the existence of exo-Oort Clouds requires the expulsion of small bodies that formed closer to their host star \citep{Oort1950}. Those small bodies could be perturbed by planets onto wide-separation orbits where ensuing stellar flybys would lift their perihelia \citep{dones2004oort}. By finding the distance at which the timescales are comparable for semimajor axis and perihelion change from planets and stellar flybys, respectively, the characteristic $10^{4}-10^{5}\,\text{au}$ heliocentric distances of the solar Oort Cloud may be derived \citep{heisler1986galacticTide}.

If planetary scattering is the primary method to generate exo--Oort Clouds, the occurrence of Jurads will depend on the prevalence of giant exoplanets. The efficiency by which (exo)planets perturb small bodies is parameterized by the Safronov number,

\begin{equation} \label{eq:safronov}
    \Theta = \frac{M_{\mathrm{p}}a_{\mathrm{p}}}{M_{*}R_{\mathrm{p}}}\,,
\end{equation}

\noindent where $M_{\text{p}}$, $R_{\text{p}}$, and $a_{\text{p}}$ are the perturbing planet's mass, radius, and semimajor axis, respectively.

\begin{figure}
    \centering
    \epsscale{1.2}
    \plotone{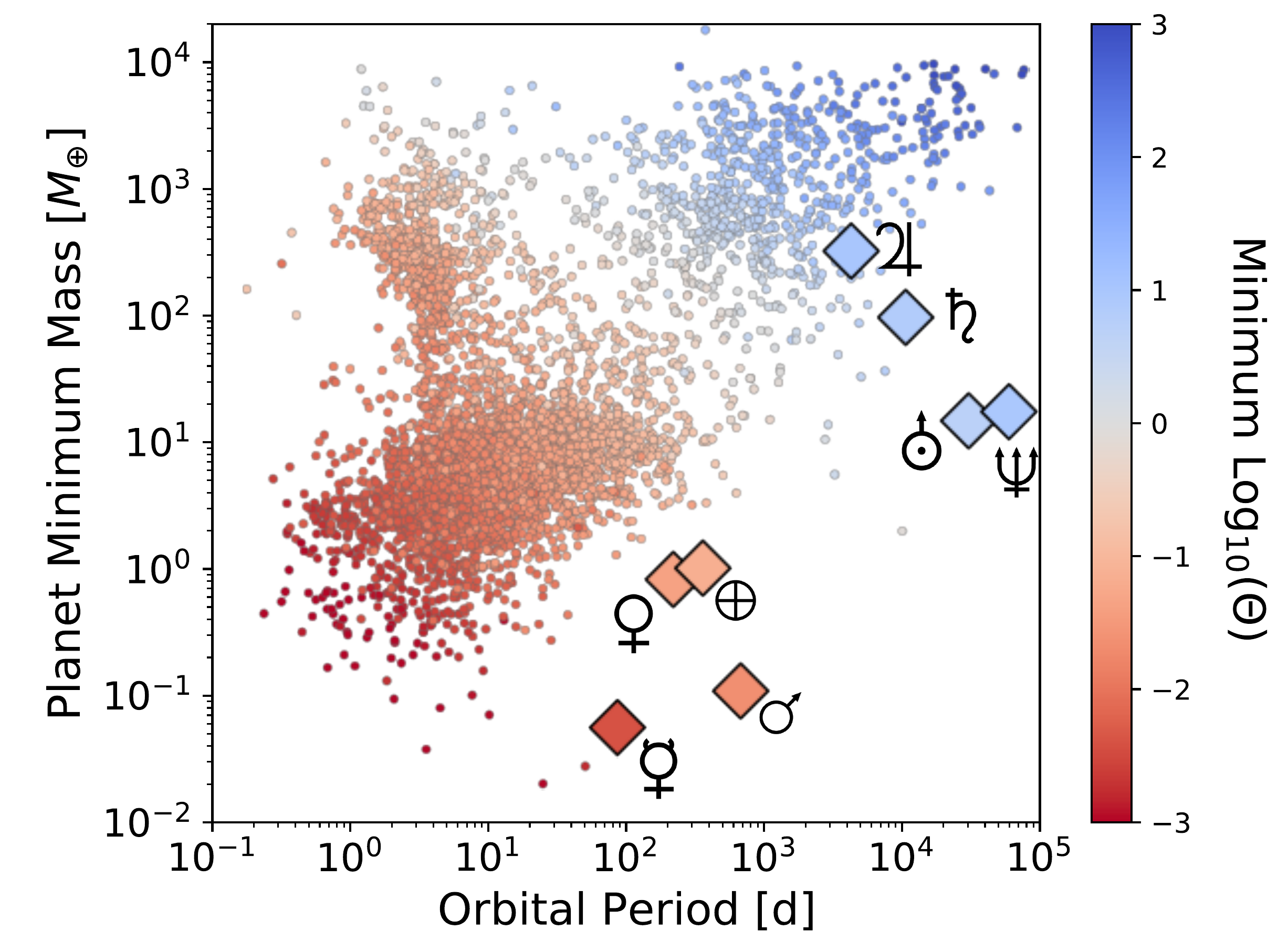}
    \caption{Masses of the known extrasolar planets (or minimum mass in the case of non-transiting planets detected by the radial velocity method) versus orbital period, where the markers have been colored by their Safronov number (Equation \ref{eq:safronov}). The solar system's planets are denoted with large diamond markers and their corresponding symbols.}
    \label{fig:safronov}
\end{figure}

Perturbers with $\Theta \gtrsim 1$ readily eject exo-comets to interstellar space. Thus, $\Theta \simeq 1$ exoplanets should accompany any Jurad-forming stars in the planet-scattering model of Oort Cloud formation. In the NASA Exoplanet Archive\footnote{Planetary Systems Composite Table (accessed January 29, 2023); DOI 10.26133/NEA13} \citep{exoplanetArchive}, approximately $12\%$ of the confirmed extrasolar planets have $\Theta > 1$ (Figure \ref{fig:safronov}). Planets in this parameter space are difficult to detect, however, as transit probabilities are low and Doppler velocities induced on the host stars are small and low frequency.

Because only a limited range of $\Theta$ values can populate an exo-Oort Cloud of small bodies, as detailed in \cite{tremaine1993pulsarOortCloud} and expanded on by \cite{wyatt2017designOortCloud}, configurations like our solar system's Oort Cloud could be rare. For the LSST to find a Jurad, the galactic ensemble of 1-8$\,\msun$ stars must (on average) generate and retain a solar-like Oort Cloud for their entire main sequence. If more complete surveys reveal a dearth of Oort Cloud-forming exoplanets, then expectations for the LSST's Jurad yield would be diminished.

Furthermore, Jurads could only form around hosts with $M_{\text{MS}} \gtrsim 1\,\msun$. Only about $6\%$ of the confirmed exoplanets jointly satisfy the $M_{\text{MS}}$ and $\Theta$ criteria, which can be attributed to the aforementioned detection biases and the dearth of usable spectral lines for radial velocity measurements of hot stars. On similar timescales to the ten-year LSST campaign, however, microlensing detections from the \textit{Roman Space Telescope} \citep{penny2019microlensingRoman} and astrometric measurements from \textit{Gaia} \citep{perryman2014astrometryGaiaExoplanet} will expand the exoplanet catalog in these regions of parameter space. Therefore, the LSST's (non)detection of Jurads will complement these space-based surveys and improve our understanding of planetary systems around intermediate mass stars.

Although Jurads are processed by post--main sequence evolution, these objects still must originally form in a circumstellar disk. Pre--main sequence evolution is rapid for higher-mass stars \citep{KWWtextbook}, so planetesimals have less time to form around potential Jurad-spawning stars. Identifying Jurads would also constrain the timing and ubiquity of planetesimal formation. Taking a $0.1\,\msun$ protosolar nebula as a reference, the solar system's Oort Cloud would have sequestered $0.3\%$ of the metal budget from the Sun's circumstellar disk. For the LSST to detect a Jurad, the exo--Oort Cloud formation efficiency must be at least this benchmark for the stars-of-interest.

We have not considered the effect of extrasolar planets on the orbital dynamics of post--main sequence environments. Encounters with giant planets with $a_{\text{p}} \gtrsim 1\,\text{au}$ could alter the ejection statistics from Section \ref{sec:ejection}, and assessing this possibility would require detailed dynamical simulations.

\subsection{Observational Clues of Jurad Origin}

Sections \ref{sec:ejection} \& \ref{sec:processing} indicated that AGB environments could alter the attributes of their exo-Oort Cloud comets and that Jurads are more likely to be ejected near their perihelia. For the same initial semimajor axis, ejected Jurads will experience more processing than their unejected counterparts.

The most telling evidence of post--main sequence processing would come from surface chemical signatures if material is deposited onto Jurads from the AGB outflows. To be observable in the solar system, however, the PAHs, dust grains, and organics would need to persist through the interstellar journey. Gas drag has been hypothesized to remove sub-micron grains from the solar system's long-period comets during passages through molecular clouds \citep{stern1990ISMerosion}, so the non-detection of these signatures cannot exclude a Jurad origin. Cosmic rays and interstellar radiation may also destroy complex molecules. Should accreted dust from the AGB outflows survive, then we predict that the Jurads from inner exo--Oort Clouds will be deeply red. If the Jurad displays a coma, then spectroscopic characterization may be feasible. Searches for these substances could be conducted at infrared wavelengths with telescopes like JWST. An intercept spacecraft \citep{Seligman2018, snodgrass2019european} would provide the most detailed species-wise abundances, but this mission would be difficult to execute due to the large expected velocity differences between Jurads and the Earth.

Our thermal modeling in Figure \ref{fig:oortCloudTemperatures} indicates that hypervolatiles like CO may be depleted from the surfaces of Jurads that originate from inner exo--Oort Clouds. Nonetheless, we caution that our chosen temperatures are not definitive values for sublimation; volatile destruction occurs over a range of temperatures and is a complicated process that depends on the surface composition, porosity, and other physical properties \citep{schorghofer2008mainBeltIce}. Hypervolatiles that would sublimate in their pure form could remain in exo-comets due to either their higher binding energies with other ices or their inability to access the surface. In the latter case, entrapment in amorphous ice \citep{jewitt2009activeCentaurs, prialnik2022amorphous} is a preservation mechanism. Since the interstellar journey cannot replenish hypervolatiles \citep{hoang2020destruction}, Jurads should retain any bulk compositional changes from thermal processing.

Moreover, devolatilization is not distinct to post--main sequence environments. Long-period comet outgassing in the solar system is usually dominated by H$_{2}$O ice even though the Sun has not yet reached the post--main sequence. Instead, Oort Cloud comets are believed to have originated from inside of the Sun's CO ice line. Scattering to the Oort Cloud is hypothesized to have occurred after the destruction of hypervolatiles, should any have existed at the small bodies' formation \citep{Lisse2022, parhi2023KBOsublimation}. Should analogous formation temperatures and scattering timescales for the creation of exo--Oort Clouds be prevalent for Jurad-spawning stars, then bulk composition may not be a strong indicator of post--main sequence processing. \cite{stern1988flyby} also showed that the top $10\,\text{m}$ of Oort Cloud objects might be heated from stellar flybys. Solar system comets exhibit a range of H$_2$O activity \citep{seligman2022borisovCO} that is modestly correlated with their size, so no single bulk composition test should set Jurads apart from exo-comets that were ejected during the host star's main sequence.

Due to the lack of definitive diagnostics, discerning whether a given interstellar small body has experienced a post--main sequence environment will rely on an assessment of inbound kinematics, detailed follow-up characterization, and a contextualization within the population of LSST-discovered interlopers. If nearly all future exo-comets resemble Borisov's observed richness of hypervolatiles and thin disk-like kinematics \citep{Cordiner2020}, then the detection of a fast-moving, CO-depleted, and deeply red interloper would be evidence in favor of a Jurad origin. The relative counts of Jurad-like to definitively non-Jurad interlopers will be important to constraining the fraction of minor planets that are ejected on the main sequence versus the post--main sequence, and this metric may also help identify a Jurad among the pool of LSST discoveries.

\subsection{The Origin of Presently-Known Interlopers}

A natural question is whether either known interstellar interloper, \om{} or Borisov, could be a Jurad. Since Pan-STARRS1 is an order-of-magnitude less sensitive to interlopers than the LSST \citep{ivezic2019lsst}, finding a Jurad in already-existing surveys is unlikely unless $M_{\text{J},*} \gtrsim 5\,\text{M}_{\oplus}/\msun$. Nonetheless, it is worthwhile to compare the physical and chemical characteristics of \om{} and Borisov to the hypothesized Jurads.

Perhaps the three most defining properties of \om{} were its extreme aspect ratio \citep{Jewitt2017,Meech2017,bannister2017col,Knight2017,drahus2018tumbling,mashchenko2019modelling}, its lack of detected volatiles \citep{Trilling2018}, and its nongravitational acceleration \citep{micheli2018non}. These attributes set \om{} apart from any solar system minor planets, although few other minor planets of \om's size have been examined with such photometric, astrometric, and spectroscopic detail. \om's observed physical properties do not prohibit a Jurad origin; the mass-wasting and thermal processing described in Section \ref{sec:processing} would be consistent with \om's shape and color \citep{Fitzsimmons2017, masiero2017spectrum, Bolin2017, bannister2017col}.

Based on our Section \ref{sec:processing} calculations, \om{} would have needed to originate from a smaller semimajor axis than an exo--Oort Cloud to match the observed axial ratio from either stellar wind processing or massive devolatilization. Indeed, some authors have proposed this history for \om{} \citep{hansen2017postMS, Rafikov2018b, katz2018interstellar} but had difficulty reconciling the non-ballistic trajectory. Should some water ice have dissociated during the interstellar journey and outgassed as H$_2$ during the solar encounter like the hypothesis by \citep{bergner2023H2}, however, then a post--main sequence origin may be feasible.

Despite the possible physical resemblance to Jurads, \om's kinematics were inconsistent with a post--main sequence origin. \om's radial and vertical inbound velocities were small versus the Local Standard of Rest (LSR), which point towards a young age \citep{Mamajek2017, gaidos2018and}. If \om{} were a Jurad, this first interloper's Galactic orbit would be an outlier among its peers. However, we note that adverse selection effects exist against interlopers with high $v_{\text{iso}, \infty}$ \citep{engelhardt2017observational, hoover2022population}.

\cite{Meech2017} and \cite{do2018interstellar} found a number density for \om-like interlopers of $0.1\,\text{au}^{-3}$, implying that typical stars generate on the order of $10\,\text{M}_{\oplus}$ of small body ejecta. \cite{do2018interstellar}'s calculation assumed \om-like kinematics of $v_{\infty} = 26\,\text{km}\,\text{s}^{-1}$. If \om{} were instead part of the Jurad population and an observational bias hampered Pan-STARRS1's ability to detect faster-moving objects, then the exo--Oort Cloud mass budget required to make \om's detection a likely event is unpalatable within modern planet-formation models.

Borisov's coma was rich in the hypervolatile CO \citep{Bodewits2020,Cordiner2020}, a substance which could be destroyed in inner exo--Oort Clouds during the post--main sequence processing of $r < 0.5\,\text{km}$ comets. Therefore, we conclude that Borisov either escaped from its host system before the star reached the AGB or resided at a large ($>10^{4}\,\text{au}$) semimajor axis before ejection during the post--main sequence. As the LSST finds more interstellar small bodies, comparing the properties of \om{} and Borisov to future interlopers will illuminate the processes that govern the formation and fate of extrasolar planetesimals.

\section{Conclusion \& Summary} \label{sec:conclusions}

The LSST's capability to discover interstellar interlopers will open novel lines-of-inquiry towards understanding extrasolar minor planets. Beginning with the observation that AGB stars should eject (Section \ref{sec:ejection}) and heat (Section \ref{sec:processing}) exo--Oort Cloud comets to the sublimation temperatures of hypervolatiles, we found that the post--main sequence environment could leave observable signatures on any Jurads that serendipitously pass through the LSST's search volume.

In this study, we have assumed fiducial values in line with the prevailing theories of (exo)planetary formation. Given the community's current understanding of Oort Clouds and exoplanet demographics, it seems unlikely that the LSST will detect a Jurad unless small body SFDs are different than those of our solar system (Sections \ref{sec:occurrence}, \ref{sec:LSST}, \& \ref{sec:constraints}). However, the lack of direct observational constraints leads to order-of-magnitude uncertainty on the total masses of exo-Oort Clouds and leaves the characteristic SFDs of exo-comets unknown.

Because these thermally processed exo-comets may be distinguishable from planetesimals ejected during the (pre--)main sequence lifetime of host stars, comparing the number of interlopers from the ``Jurad" pathway those of other classes could be possible. This tabulation would probe the occurrence of exo--Oort Clouds, thereby providing insight into the ubiquity of wide-separation giant planets and their potential migratory histories.

\vspace{0.5cm}
ACKNOWLEDGEMENTS: We would like to express our gratitude to both referees for their detailed and insightful reviews which improved the scientific content of this study. We thank Juliette Becker, David Hernandez, Tiger Lu, Yubo Su, Chris O'Connor, and Emma Louden for useful discussions. We are grateful to Christopher Lindsay for pointing us towards resources on standardized stellar evolutionary tracks. WGL acknowledges support from the Department of Defense's National Defense Science \& Engineering Graduate (NDSEG) Fellowship. DZS acknowledges financial support from the National Science Foundation  Grant No. AST-2107796, NASA Grant No. 80NSSC19K0444 and NASA Contract  NNX17AL71A from the NASA Goddard Spaceflight Center. This research has made use of the NASA Exoplanet Archive, which is operated by the California Institute of Technology, under contract with the National Aeronautics and Space Administration (NASA) under the Exoplanet Exploration Program.

\software{\texttt{numpy} \citep{harris2020numpy}, \texttt{scipy} \citep{virtanen2020scipy}, \texttt{matplotlib} \citep{hunter2007matplotlib}.}

\bibliography{bibliography}
\bibliographystyle{aasjournal}

\appendix

\section{Dynamical Model Validation} \label{sec:dynamicalModelValidation}

To validate our numerical post--main sequence dynamical model (Figure \ref{fig:ejectionStats}), we initialized the \texttt{rebound} simulation that was described in Section \ref{sec:ejection} with $3\times10^{3}$ test particles. We performed the integration five times, keeping the same initial conditions for the set of test particles but varying the discrete timestep over which we modified the stellar mass: $t_{s} = \{1, 10, 100, 1000, 2500\}\,\text{yr}$. In each run, timesteps for dynamical evolution are chosen adaptively by the internal IAS15 integrator. The orbital timestamps are not constant between simulations or within a simulation itself.

\begin{figure}
    \centering
    \includegraphics[width=\linewidth,angle=0]{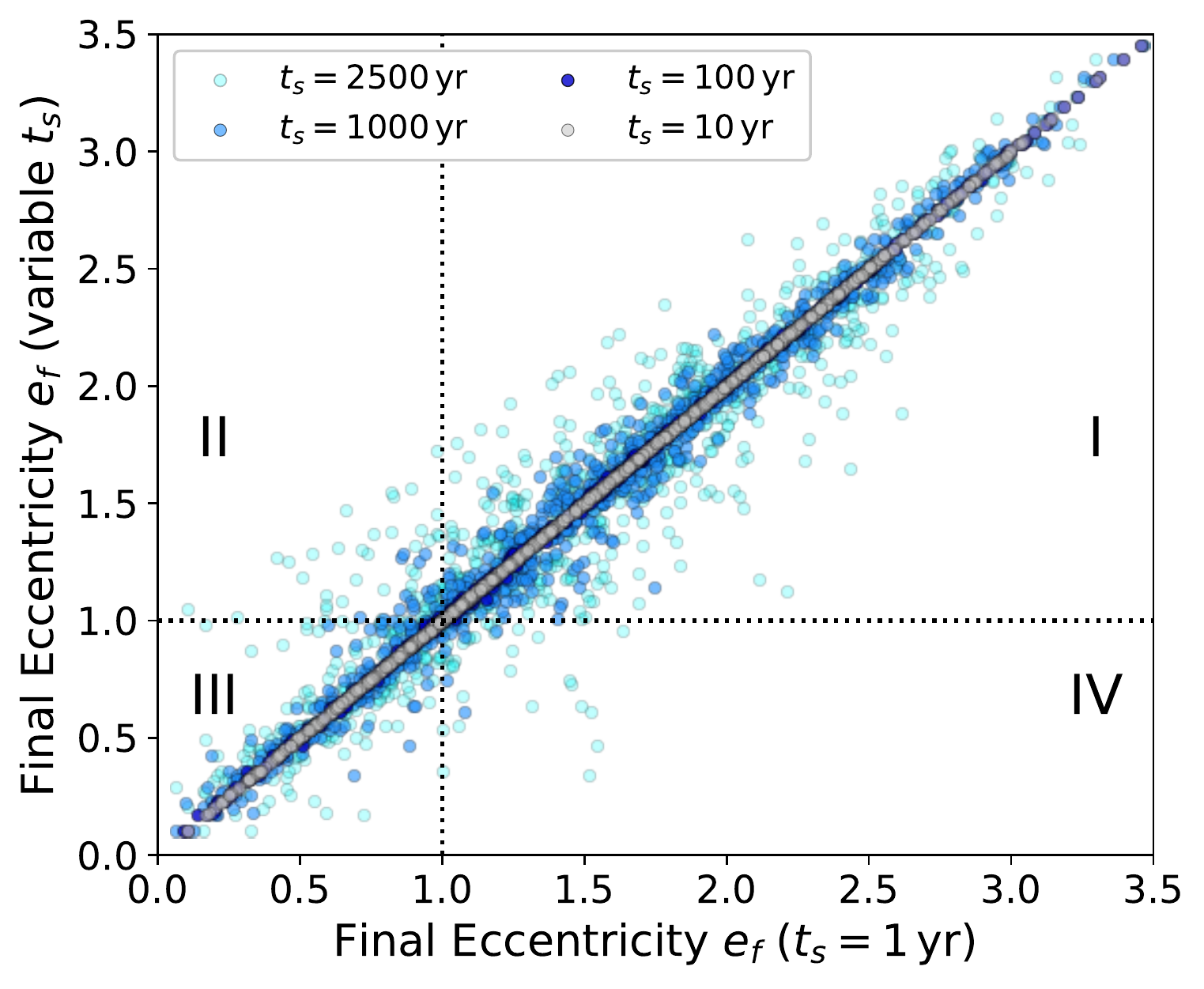}
    \caption{Object-by-object comparison of final eccentricities $e_{\text{J}, f}$ for test particles in the validation \texttt{rebound} simulations that were run with different mass loss timestamps $t_{s}$. The y-axis shows $e_{\text{J}, f}$ for $t_{s} = \{10, 100, 1000, 2500\}\,\text{yr}$ versus $e_{\text{J}, f}$ for the $t_{s} = 1\,\text{yr}$ simulation on the x-axis. Quadrants are labeled I-IV, with regions II and IV representing parameter space where particles are bound ($e<1$) in one simulation and unbound ($e>1$) for the other simulation.}
    \label{fig:validateRebound}
\end{figure}

We check the consistency of the final eccentricities $e_{\text{J}, f}$ for runs with the aforementioned mass loss timestamps. Values of $t_{s}$ that are numerically stable should return constant values of $e_{\text{J}, f}$ for a given object. Figure \ref{fig:validateRebound} shows the final eccentricities for $t_{s} = \{10, 100, 1000, 2500\}\,\text{yr}$ versus the values that resulted from the $t_{s} = 1\,\text{yr}$ run. The distribution of $e_{\text{J}, f}$ from the the $t_{s} = 100\,\text{yr}$ timestamps is close to that from the $t_{s} = 10\,\text{yr}$ distribution. Moreover, both of these sets of $e_{\text{J}, f}$ are nearly identical with the set from $t_{s} = 1\,\text{yr}$. In total, Figure \ref{fig:validateRebound} shows that $t_{s} = 10\,\text{yr}$ timestamps are appropriate.

Although Figure \ref{fig:validateRebound} demonstrates the numerical stability of our \texttt{rebound} simulations for $t_{s} = 100\,\text{yr}$, some errors in $e_{\text{J}, f}$ would not affect the results in Section \ref{sec:ejection}. Our final results would only change if objects that should remain bound are ejected or vice versa for two different timesteps $t_{s}$. These cases correspond to points in quadrants II and IV on Figure \ref{fig:validateRebound}. We report the fraction of points for each $t_{s}$ that fall in these quadrants on Table \ref{tab:validateRebound}. Although we could likely take $t_{s} > 100\,\text{yr}$ while maintaining the fidelity of our Section \ref{sec:ejection} results, we elect to use this conservative timestamp based on the results of Figure \ref{fig:validateRebound} and Table \ref{tab:validateRebound}.

\begin{table}[]
\centering
\caption{Fraction of test particles in validation simulations with $e_{f}$ errors versus the $t_{s} = 1$ simulation that flip the final status of the exo-comet from either bound-to-unbound (Quadrant II) or unbound-to-bound (Quadrant IV).}
\label{tab:validateRebound}
\begin{tabular}{|c|c|c|}
\hline
\textbf{$t_{s}$ {[}yr{]}} & \textbf{Quadrant II} & \textbf{Quadrant IV} \\ \hline
10 & 0.000 & 0.000\\ \hline
100 & 0.001 & 0.001\\ \hline
1000 & 0.007 & 0.011\\ \hline
2500 & 0.018 & 0.024\\ \hline
\end{tabular}
\end{table}

As another check for the appropriateness of $t_{s} = 100\,\text{yr}$, we calculated the degree to which the  specific angular momentum $h_{\text{J}}$ was conserved. This quantity is an integral of motion \citep{hadjidemetriou1963variableMass} for isotropic mass loss and should be conserved in our numerical simulations. We checked the value of $h$ as reported by $\texttt{rebound}$ and as calculated directly from the position-velocity values $(x, y, v_x, v_y)$. For all test particles in the $t_{s} = 100\,\text{yr}$ run, we find that the fractional change of the conserved quantity $h$ is always less than $9\times10^{-15}$ with both methods.

\section{Thermal Model Validation}\label{sec:thermmodelvalid}

Here, we provide validation tests for the numerical accuracy of the 1-D thermal model used and presented in Section \ref{subsec:thermprocess}. This model involves both a non-linear surface radiation term and spherical geometry. Therefore, validation via numerical evaluation is nontrivial, even for analytic solutions. We use Mathematica \citep{Mathematica} to compute high-accuracy numerical simulations for comparison. While the Mathematica simulations are highly accurate, they are not suitable to the analysis of thermal penetration over these time scales, due to efficiency issues. Our own model implementation allows for the convenient use of stellar luminosities on the boundary conditions, easy parallelization of model instances, and is significantly more efficient --- at relevant time scales. The wall time for our implementation is $t\sim5$ seconds, versus $t\sim5$ minutes for Mathematica, a $60\times$ improvement.  

To produce the Mathematica simulations, we use the \texttt{NDSolve} method to solve  the  heat equation on a sphere (Equation \ref{eq:heat}). To address the central pole, we set the innermost point to be at $10^{-14}$ cm. We verified that variations in this value do not notably change our results. At this central point, we impose a symmetric boundary condition, and at the surface, we impose a radiative boundary condition with an external blackbody temperature of 30 K. Both of these conditions are imposed using  built-in boundary condition functions in Mathematica. For the thermal parameters, we set the albedo to be $\mathcal{A}=0.1$, the thermal diffusivity $k=10^3$ ergs cm$^{-1}$ s$^{-1}$, the heat capacity $c_P=2\cdot10^7$ ergs g$^{-1}$ K$^{-1}$, and the density $\rho=0.5$ g cm$^{-3}$. The simulation was initialized at 3 K and run for $10^{12}$ seconds ($3\cdot10^5$ years). The temperature is saved at a grid of depths and times for comparison to our model.

To produce equivalent results with our model, we use identical initial condition and thermal parameters. Because the radiative boundary condition in Mathematica specifies a background temperature (rather than a heat flux), we set the orbital distance $a=100$ au, and the stellar luminosity to a constant $L_*=1.346\lsun$. This value corresponds to an effective temperature of 30 K, identical to the Mathematica simulation.  We use a time step of $\Delta t=10^7$ seconds and a radial discretization of $\Delta r=75$ cm. We  compute the temperature over the same grid of depths and times.

\begin{figure}
    \centering
    \includegraphics[width=\linewidth,angle=0]{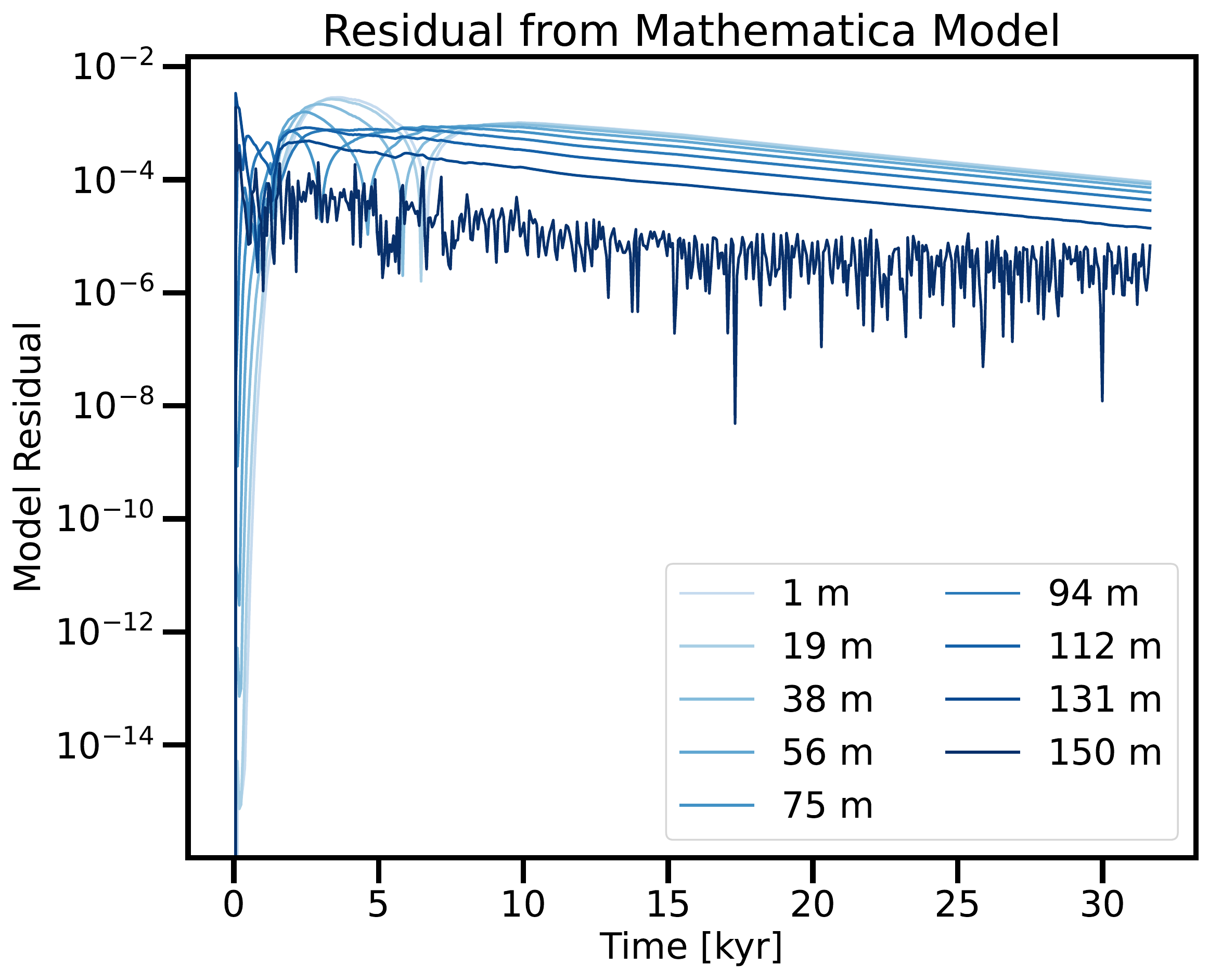}
    \caption{Residual of the thermal model presented in this paper with regards to a similar model constructed in Mathematica.}
    \label{fig:thermmodelerror}
\end{figure}

Both simulations produce qualitatively and quantitatively similar results. The residual, computed as $\|T_{\rm model}-T_{\rm Math.}\|/T_{\rm Math}$, is shown in Figure \ref{fig:thermmodelerror}. The residual attains a maximum value of $2.9\cdot10^{-2}$ and a median value of $2.16\cdot10^{-4}$. The maximum value is obtained at the very beginning of the simulation, where the solution relaxes to a stable configuration. This error is convergent, and is independent of the time step, the spatial discretization, or the simulation time. Given that our results are not sensitive to temperature variations on the order of fractions of a degree Kelvin, our numerical method is sufficient. 

\end{document}